\documentclass{emulateapj}

\newcommand{\lya}{Ly$\alpha$}
\newcommand{\ha}{H$\alpha$}
\newcommand{\hb}{H$\beta$}
\newcommand{\oiii}{[O\,{\sc iii}] $\lambda$5007}
\newcommand{\heii}{He\,{\sc ii} $\lambda$1640}

\begin{document}

\title{PHYSICAL PROPERTIES OF SPECTROSCOPICALLY-CONFIRMED GALAXIES AT $z\ge6$. 
I. BASIC CHARACTERISTICS OF THE REST-FRAME UV CONTINUUM AND LYMAN-$\alpha$ 
EMISSION\footnotemark[$\ast$]}
\footnotetext[$\ast$]{
Based in part on observations made with the NASA/ESA Hubble Space Telescope, 
obtained from the data archive at the Space Telescope Science Institute, which 
is operated by the Association of Universities for Research in Astronomy, Inc. 
under NASA contract NAS 5-26555.
Based in part on observations made with the Spitzer Space Telescope, which is 
operated by the Jet Propulsion Laboratory, California Institute of Technology 
under a contract with NASA.
Based in part on data collected at Subaru Telescope and obtained from the
SMOKA, which is operated by the Astronomy Data Center, National Astronomical
Observatory of Japan.}

\author{Linhua Jiang\altaffilmark{1,2,3}, Eiichi Egami\altaffilmark{2},
Matthew Mechtley\altaffilmark{1}, Xiaohui Fan\altaffilmark{2}, 
Seth H. Cohen\altaffilmark{1}, Rogier A. Windhorst\altaffilmark{1},
Romeel Dav\'{e}\altaffilmark{2,4}, Kristian Finlator\altaffilmark{5},
Nobunari Kashikawa\altaffilmark{6}, Masami Ouchi\altaffilmark{7,8},
and Kazuhiro Shimasaku\altaffilmark{9}}
\altaffiltext{1}{School of Earth and Space Exploration, Arizona State 
	University, Tempe, AZ 85287-1504, USA; linhua.jiang@asu.edu}
\altaffiltext{2}{Steward Observatory, University of Arizona,
   933 North Cherry Avenue, Tucson, AZ 85721, USA}
\altaffiltext{3}{Hubble Fellow}
\altaffiltext{4}{Physics Department, University of the Western Cape, 7535 
	Bellville, Cape Town, South Africa}
\altaffiltext{5}{DARK fellow, Dark Cosmology Centre, Niels Bohr Institute, University of Copenhagen}
\altaffiltext{6}{Optical and Infrared Astronomy Division, National
   Astronomical Observatory, Mitaka, Tokyo 181-8588, Japan}
\altaffiltext{7}{Institute for Cosmic Ray Research, The University of Tokyo,
	5-1-5 Kashiwanoha, Kashiwa, Chiba 277-8582, Japan}
\altaffiltext{8}{Kavli Institute for the Physics and Mathematics of the 
	Universe, The University of Tokyo, 5-1-5 Kashiwanoha, Kashiwa, Chiba 
	277-8583, Japan}
\altaffiltext{9}{Department of Astronomy, University of Tokyo, Hongo, Tokyo 
	113-0033, Japan}

\begin{abstract}

We present deep {\it HST} near-IR and {\it Spitzer} mid-IR observations of a 
large sample of spectroscopically-confirmed galaxies at $z\ge6$. The sample
consists of 51 \lya\ emitters (LAEs) at $z\simeq5.7$, 6.5, and 7.0, and 16
Lyman-break galaxies (LBGs) at $5.9\le z\le6.5$. The near-IR images were 
mostly obtained with WFC3 in the F125W and F160W bands, and the mid-IR images
were obtained with IRAC in the $3.6\mu$m and $4.5\mu$m bands. Our galaxies 
also have deep optical imaging data from Subaru Suprime-Cam. We utilize the 
multi-band data and secure redshifts to derive their rest-frame UV properties.
These galaxies have steep UV continuum slopes roughly between 
$\beta\simeq-1.5$ and --3.5, with an average value of $\beta\simeq-2.3$, 
slightly steeper than the slopes of LBGs in previous studies. 
The slope shows little dependence on UV continuum luminosity except 
for a few of the brightest galaxies. We find a statistically significant 
excess of galaxies with slopes around $\beta\simeq-3$, suggesting the 
existence of very young stellar populations with extremely low metallicity and 
dust content. Our galaxies have moderately strong rest-frame \lya\ equivalent 
width (EW) in a range of $\sim$10 to $\sim$200 \AA. The star-formation rates 
are also moderate, from a few to a few tens solar masses per year. 
The LAEs and LBGs in this sample share many common properties, implying that
LAEs represent a subset of LBGs with strong \lya\ emission.
Finally, the comparison of the UV luminosity functions between LAEs and LBGs
suggests that there exists a substantial population of faint galaxies with 
weak \lya\ emission ($\rm EW<20$ \AA) that could be the dominant contribution 
to the total ionizing flux at $z\ge6$.

\end{abstract}


\section{INTRODUCTION}

The epoch of cosmic reionization marks one of the major phase transitions of
the Universe, during which the neutral intergalactic medium (IGM) was ionized 
by the first astrophysical objects. Measurements of CMB polarization have 
shown that reionization began earlier than $z\simeq10.6$ \citep{kom11}. 
Meanwhile, studies of Gunn-Peterson troughs in high-redshift quasar spectra 
have located the end of reionization at $z\simeq6$ \citep{fan06}. Therefore, 
objects at $z\ge6$ are natural tools to probe this epoch. With recent advances 
of instrumentation on {\it Hubble Space Telescope} ($HST$) and large 
ground-based telescopes, galaxies at $z\ge6$ are now being routinely found. 
Direct observations of the earliest galaxy formation and the history of 
reionization are finally in sight \citep{rob10,finlator12}.

The first $z>6$ galaxies were discovered to be \lya\ emitters (LAEs) at 
$z\simeq6.5$ using the narrow-band (or \lya) technique 
\citep{hu02,kod03,rho04}. 
This technique has been an efficient way to find high-redshift galaxies due to 
its high success rate of spectroscopic confirmation. Three dark atmospheric 
windows with little OH sky emission in the optical are often used to detect
distant galaxies at $z\simeq5.7$, 6.5, and 7. So far more than 200 LAEs have 
been spectroscopically confirmed at these redshifts 
\citep[e.g.,][]{tan05,iye06,shi06,kas06,kas11,hu10,ouc10,rho12}. 
The narrow-band technique is also being used to search for higher redshift 
LAEs at $z>7$ \citep[e.g.,][]{hib10,til10,cle12,kru12,ota12,shibuya12}. 
All these \lya\ surveys were made with ground-based instruments owing to 
their large field-of-views (FOVs) and relatively low sky-background in 
OH-dark windows.

A complementary way to find high-redshift galaxies is the dropout technique 
\citep{ste93,gia02}. It has produced a large number of Lyman-break galaxies 
(LBGs) or candidates at $z\ge6$. While large-area ground-based observations 
are efficient to select bright LBGs 
\citep[e.g.,][]{nag07,bow12,cur12,hsi12,wil13}, 
the majority of the faint LBGs at $z\ge6$ known so far were discovered by 
$HST$ \citep[e.g.,][]{bun04,dic04,yan04}.
In particular, with the power of the new $HST$ WFC3 infrared camera, a few 
hundred galaxies at $6\le z\le12$ have been detected recently 
\citep[e.g.,][]{bun10,fin10,mcl10,oes10,wil10,yan10,bou11,lor11,ellis13}. 
Although most of the LBG candidates are too faint for follow-up identification,
some of them have been spectroscopically confirmed with deep 
ground-based observations
\citep[e.g.,][]{sta10,sta11,jia11,van11,pen11,ono12,schenker12}.

With the large samples of galaxies (and candidates) at $z\ge6$, the UV 
continuum luminosity function (LF) and \lya\ LF are being established 
\citep[e.g.,][]{hu10,jia11,kas11,bra12,bou12a,hen12,lap12,oes12}. 
It is found that the faint-end slope of the UV LFs are very steep 
($\alpha\simeq-2$), so dwarf galaxies could provide enough UV photons for 
reionization (also depending on the ionizing photon escape fraction and the 
ionized gas clumping factor). For individual galaxies, their physical 
properties are also being investigated. At $z\ge6$, the rest-frame UV/optical 
light moves to the infrared range. Therefore, infrared observations, especially 
the combination of $HST$ and {\it Spitzer Space 
Telescope} ($Spitzer$), are essential to measure properties of these 
high-redshift galaxies. $HST$ near-IR data constrain the slope of the 
rest-frame UV spectrum and help to decipher the properties of young stellar 
populations. Recent $HST$ WFC3 near-IR observations have shown that 
high-redshift LBGs have blue rest-frame UV slopes $\beta$
($f_{\lambda} \propto {\lambda}^{\beta}$). The typical values of $\beta$ are
close to or smaller (bluer) than --2 
\citep[e.g.,][]{wil11,bou12b,dun12a,fin12}. 
Though the selection criteria of LBGs and the approaches to UV slopes are 
slightly different in the different studies in the literature, it is believed 
that $z\ge6$ galaxies are bluer than low-redshift galaxies, implying
that they are generally younger and have low dust extinction. 

More detailed physical properties (e.g. age and stellar mass) of high-redshift 
galaxies have come from SED modeling based on broad-band photometry of 
$HST$ and $Spitzer$ as well as optical data. $Spitzer$ IRAC provides 
mid-IR photometry. When combined with $HST$ near-IR data, it measures the 
amplitude of the Balmer 
break and constrains the properties of mature populations. The early results 
on $z\simeq6$ LBGs showed IRAC detections, and emphasized galaxies with 
established stellar populations \citep[e.g.,][]{ega05,eyl05,yan05}. Later 
studies found that most LBGs were actually not detected 
in moderately deep IRAC images, 
suggesting that these galaxies were considerably younger and less massive 
\citep[e.g.,][]{yan06,eyl07}. LAEs may represent an even more extreme 
population with ages of only a few million years \citep[e.g.,][]{pir07}.
As the number of galaxies known at $z\ge6$ increases steadily, extensive
studies are being carried out on various galaxy samples 
\citep[e.g.,][]{gon10,ono10,sch10,mcl11,pir12}. A diversity of physical
properties are found among these galaxies, although the Universe is younger
than one billion years.

Despite the progress that has been made on properties of high-redshift 
galaxies, there is a lack of a large, reliable sample of $z\ge6$ galaxies with
spectroscopic redshifts and high-quality infrared photometry. The reason is 
twofold. For galaxies found by ground-based telescopes --- although they are 
relatively bright and have spectroscopic redshifts --- they often do not have 
measured infrared (rest-frame UV/optical) information. For example, only a
small amount of LAEs at $z\ge6$ have been observed in the near-IR bands by 
$HST$ \citep[e.g.,][]{cowie11a}. On the other hand, 
galaxies found by $HST$ (e.g. those in the GOODS fields), do have near-IR 
photometry, but most of them are too faint for follow-up spectroscopic 
identification by current facilities. So they do not have secure redshifts, 
making it difficult for accurate SED modeling. A model SED derived from a few 
photometric points without a secure redshift is usually highly degenerate. 
Therefore, determination of the physical properties of $z\ge6$ galaxies 
demands a large sample of galaxies with both spectroscopic redshifts 
and high-quality infrared photometry.

In this paper we will present deep $HST$ and $Spitzer$ observations of 67
spectroscopically-confirmed (hereafter spec-confirmed) galaxies at $z\ge6$. 
This galaxy sample contains 51 LAEs and 16 LBGs, and is the largest collection 
of spec-confirmed galaxies in this redshift range. The spectroscopic redshifts 
of the sample provide great advantages for measuring the physical properties 
of galaxies. First, this sample is uncontaminated by interlopers. 
A photometrically-selected (hereafter photo-selected) sample of high-redshift 
galaxies may be contaminated by interlopers such as low-redshift red or dusty 
galaxies and Galactic late-type dwarf stars, which may cause various bias on 
measurements of physical properties. 
Second, spectroscopic redshifts remove one critical free parameter for SED 
modeling. For the brightest galaxies at $z\ge6$, there are usually only 4--5 
broad-band photometric points available 
(e.g., two $HST$ bands, two $Spitzer$ IRAC bands, and one optical band) for 
producing stellar population models. Given the limited degrees of freedom, 
a spectroscopic redshift will significantly improve SED modeling,
especially when nebular emission is considered. At $z\ge6$, strong nebular
emission lines such as \oiii, \hb, and \ha\ enter IRAC 1 and 2 channels, which 
affect our estimate of stellar populations 
\citep[e.g.,][]{sch10,finlator11,deb12,sta13}. Photometric redshifts with 
large uncertainties may place these nebular lines in wrong bands during SED 
fitting. With spectroscopic redshifts, the positions of nebular lines are 
secured (see Stark et al. 2012 for more discussion).
Finally, we currently have little knowledge of stellar populations in 
$z\ge5.5$ LAEs. This is because almost all the known LAEs were discovered by
ground-based telescopes, and most of them do not have infrared observations. 
Our paper includes a large sample of LAEs. This allows us, for the first time, 
to systematically study stellar populations in LAEs.

This paper is the first in a series presenting the physical properties of 
these galaxies based on observations from $HST$, $Spitzer$, and the Subaru 
telescope. The vast majority of the galaxies were discovered in the Subaru 
Deep Field \citep[SDF;][]{kas04}. The SDF is a unique and ideal field to study 
the distant Universe: it covers an area of over 800 arcmin$^2$ and has the 
deepest optical images among all ground-based imaging data. The SDF project 
has been highly successful in searching for high-redshift galaxies. It has 
identified more than 100 LAEs and LBGs at $z\ge6$ 
\citep[e.g.,][]{tan05,iye06,shi06,nag07,ota08,jia11,kas11,tos12}.
These galaxies are spectroscopically confirmed and represent the most luminous 
galaxies in this redshift range. In this paper we present the $HST$ and 
$Spitzer$ observations, and use these data together with the optical 
data from Subaru to derive the rest-frame UV properties, such as their UV 
continuum slopes $\beta$ and \lya\ luminosities. In the subsequent papers of 
this series we will present rest-frame UV morphology and measure stellar 
populations for these galaxies.

The structure of the paper is as follows. In Section 2, we present our galaxy
sample. We also describe our optical/infrared observations and data reduction.
In Section 3, we derive the rest-frame UV continuum properties of the 
galaxies. In Section 4, we measure the properties of \lya\ emission lines.
We discuss our results in Section 5 and summarize the paper in Section 6.
Throughout the paper we use a $\Lambda$-dominated flat cosmology with $H_0=70$
km s$^{-1}$ Mpc$^{-1}$, $\Omega_{m}=0.3$, and $\Omega_{\Lambda}=0.7$ 
\citep{kom11}. All magnitudes are on the AB system \citep{oke83}.

\section{OBSERVATIONS AND DATA REDUCTION}

\subsection{Galaxy Sample}

Table 1 shows the list of the galaxies presented in this paper. There are a 
total of 67 spec-confirmed galaxies at $5.6\le z\le7$: 62 of them 
are from the SDF and the remaining 5 are from the Subaru XMM-Newton Deep 
Survey \citep[SXDS;][]{fur08} field. They represent the most luminous galaxies 
in terms of \lya\ luminosity for LAEs or UV continuum luminosity for LBGs
in this redshift range.

The SDF covers an area of $\sim876$ arcmin$^2$, and its optical imaging data 
were taken with Subaru Suprime-Cam in a series of broad and narrow bands 
\citep{kas04}. Especially noteworthy are the deep observations with three 
narrow-band (NB) filters, NB816, NB921, and NB973, corresponding to the 
detection of LAEs at $z\simeq5.7$, 6.5, and 7. The full widths at half maximum 
(FWHM) of the three filters are 120, 132, and 200 \AA, respectively. 
So far the SDF 
project has spectroscopically confirmed more than 100 LAEs at $z\simeq5.7$, 
6.5, 7, and more than 40 LBGs at $6\le z \le7$. Our SDF galaxy sample contains 
22 LAEs at $z\simeq5.7$ \citep{shi06,kas11}, 25 LAEs at $z\simeq6.5$ 
\citep{tan05,kas06,kas11}, and the $z=6.96$ LAE \citep{iye06}. The LAEs at 
$z\simeq5.7$ and 6.5 were selected in a similar way, and have a relatively 
uniform magnitude limit of 26 mag in the narrow bands NB816 and NB921, so
they are from a well-defined flux-limited sample. Our SDF sample also 
contains 14 LBGs at $5.9\le z\le6.5$ \citep{nag04,nag05,nag07,ota08,jia11};
two of them have not been previously published.
These LBGs were selected with different criteria and thus have 
inhomogeneous depth. The $z'$-band magnitude limit is 26.1 mag in Nagao 
et al. papers, 26.5 mag in \citet{ota08}, and 27 mag in \citet{jia11}.
For the details of candidate selection and follow-up spectroscopy, see the
original galaxy discovery papers above.

Five galaxies in our sample, including 2 LBGs at $z\simeq6$ \citep{cur12} and 
3 LAEs at $z\simeq6.5$ \citep{ouc10}, are from the SXDS. The two LBGs are very
bright ($z'<25$ mag) and have been analyzed by \citet{cur13}. The SXDS 
optical imaging data were also taken with Subaru Suprime-Cam in the same
broad and narrow bands, but cover five times larger area than the SDF does
\citep{fur08}.

In Table 1, the first 62 galaxies are from the SDF and the last
5 galaxies are from the SXDS. Columns 2 and 3 list the object coordinates. 
They are re-calculated from our own stacked optical images (see Section 2.2),
and are fully consistent with those given in the galaxy discovery papers 
(typical difference $<0\farcs1$). Column
4 lists the redshifts, measured from the \lya\ emission lines of the galaxies.
The galaxy discovery papers used different values for the rest-frame \lya\
wavelengths (i.e. 1215, 1215.67, or 1216 \AA) for redshift measurements.
We convert all these redshifts to the redshifts listed in Table 1
by assuming a rest-frame \lya\ line center of 1215.67 \AA.
The last column shows the reference papers of the objects.
We will explain the rest of the table in the following subsections. 

\subsection{Optical Imaging Data}

The galaxies in this paper were discovered in the SDF and the SXDS. The SDF 
public imaging data include stacked images in five broad bands $BVRi'z'$ and 
two narrow bands NB816 and NB921. The flux limits for these data are given in 
\citet{kas04}. However, the public data do not include the data taken recently
(also by Subaru Suprime-Cam), and the data taken in the $y$ and NB973 bands.
Therefore, we produced our own stacked images by including all available 
images in the Subaru archive, as explained below.

Our data processing began with the raw images obtained from the archival 
server SMOKA \citep{bab02}. Raw images with point spread function (PSF) sizes 
greater than $1\farcs2$ were rejected. The data were reduced, re-sampled, and 
co-added using a combination of the Suprime-Cam Deep Field REDuction package 
\citep[{\tt SDFRED};][]{yag02,ouc04} and our own {\tt IDL} routines. Each 
image was bias (overscan) corrected and flat-fielded using {\tt SDFRED}. Then 
bad pixel masks were created from flat-field images. Cosmic rays were 
identified and interpolated using {\tt L.A. Cosmic} \citep{van01}. Saturated 
pixels and bleeding trails were also identified and interpolated. For each 
image, a weight mask was generated to include all above mentioned defective 
pixels. We then used {\tt SDFRED} to correct the image distortion, subtract 
the sky-background, and mask out the pixels affected by the Auto-Guider probe.
After individual images were processed, we extracted sources using 
{\tt SExtractor} \citep{ber96} and calculated astrometric and photometric 
solutions using {\tt SCAMP} \citep{ber06} for the final image co-addition. In 
order to match the public SDF images, we used the public $R$-band image as the 
astrometric reference catalog in {\tt SCAMP}. Both science and weight-map 
images were scaled and updated using the astrometric and photometric solutions 
measured from {\tt SCAMP}. Instead of homogenizing PSFs to a certain value
as was done for the public data, we incorporated PSF information into the 
weight image so that the weight was proportional to the inverse of the square 
of the PSF FWHM. We re-sampled and co-added images
using {\tt SWARP} \citep{ber02}. The re-sampling interpolations for science 
and weight images were LANCZOS3 and BILINEAR, respectively. We modified 
{\tt SWARP} so that it performs sigma-clipping ($5\sigma$ rejection) of
outliers when co-adding images.

We produced images in the six broad bands $BVRi'z'y$ and three narrow bands 
NB816, NB921, and NB973. The final products are a stacked science image, and 
its corresponding weight image, for each band. The pixel and image sizes of 
our products are the same as those of the public SDF images. The typical PSF 
FWHM is $0\farcs6-0\farcs7$. We determined photometric zero points from the 
public SDF images whose zero points were known. 
The SXDS optical imaging data were reduced 
in the same way as we did for the SDF.

Our SDF broad-band images are deeper than the public images in the $Ri'z'$ 
bands, because the SDF team obtained more imaging data from Subaru recently, 
and we included all the available data above. For example, the public 
$z'$-band image has a depth of $z'=26.1$ mag ($5\sigma$ detection in a 
$2\arcsec$ diameter aperture) with a total integration time of $\sim$7 hr. 
To date, the total $z'$-band data amount to $\sim$29 hr. Our 
new $z'$-band image has a depth of $\sim$27.0 mag, 0.9 mag 
deeper than the public image. In addition, the public data do not have images 
in the $y$ and NB973 bands. The SDF $y$-band imaging data were taken with two 
different sets of CCDs (MIT and Hamamatsu), and have been used to search for 
$z\simeq7$ LBGs \citep{ouc09}. Although the fully-depleted Hamamatsu CCDs are 
twice as sensitive as the MIT CCDs in the $y$ band, their sensitivity (quantum 
efficiency) curves have similar shapes. Therefore, we stacked all the images 
taken with both sets of CCDs, as \citet{ouc09} did in their work. The total 
integration time was 24 hr, and the depth was about 26.0 mag.

In Table 1, Columns 5 to 7 show the total magnitudes of the galaxies in one of
the three narrow bands (NB816, NB921, or NB973, depending on redshift) and two
broad bands $z'y$. For a given object in a given filter, its total magnitude 
is measured as follows. We first use {\tt SExtractor} to measure the aperture 
photometry of the object in a circular $2\farcs0$ diameter, after the local 
sky-background is subtracted. In cases where galaxies are close to their 
neighbors, we use a circular $1\farcs6$ diameter. Then an aperture correction 
is applied to correct for light loss. The aperture correction is determined 
from more than 100 bright, but unsaturated point sources in the same image. 
Almost all our galaxies are point-like objects and are not resolved in the SDF 
images. In cases where galaxies clearly show extended features, we use 
MAG\_AUTO magnitudes as the best estimates of the total magnitudes. In Table 
1, we did not list broad-band photometry for the galaxies with weak detections 
($<5\sigma$) in the $J$ band. These galaxies are excluded in most of our
analysis in this paper. Detections below $3\sigma$ in any of these bands are 
not listed as well. We carefully check each object, 
and find that object no. 12, a $z\simeq5.7$ LAE, is overlapped by a bright 
foreground star (this was also pointed out in the galaxy discovery paper). 
The separation between the two objects 
is only $\sim0\farcs25$, so we are not able to do photometry for this 
galaxy. We will not use this galaxy in the following analysis.

Figure 1 shows the transmission curves of six Subaru Suprime-Cam filters, 
including three broad-band filters $i'z'y$ and three narrow-band filters 
NB816, NB921, and NB973. We do not show the other three bluer filters $BVR$,
because our galaxies were not detected in these bands.
In Appendix A, we show the thumbnail images of the galaxies in one of the 
three Subaru Suprime-Cam narrow bands 
and the three broad bands $i$, $z'$, and $y$.

\subsection{$HST$ Near-IR Imaging Data}

We obtained $HST$ near-IR imaging data for a large sample of spectroscopically 
confirmed SDF galaxies at $5.6\le z\le7$ from three $HST$ GO programs 11149 
(PI: E. Egami), 12329 and 12616 (PI: L. Jiang). The three programs together 
with our $Spitzer$ programs (see the next subsection) were designed to 
characterize the physical properties of these high-redshift galaxies. 
The $HST$ observations were made with a mix of strategies. As the proposal for
GO program 11149 was written, 
$HST$ WFC3 had not yet been installed. Our plan was to observe 20 bright 
galaxies with NICMOS in the F110W (hereafter $J_{110}$) 
and F160W (hereafter $H_{160}$) bands. The observations were made 
with camera NIC3. The exposure time was two orbits per band per pointing (or 
per object), split into four exposures. The total on-source integration per 
band was about 5400 sec. We performed sub-pixel dithering during the four 
NICMOS exposures to improve the PSF sampling and to remove bad pixels 
and cosmic rays.

After we finished several targets, NICMOS failed, and then WFC3 was installed 
in May 2009. Since WFC3 has much higher sensitivity, the rest of the galaxies 
in GO 11149 and all the galaxies in GO 12329 and 12616 were observed with WFC3. 
Because of the larger FOV of WFC3 and the increasing number of known $z\ge6$ 
galaxies in the SDF, we changed our observing strategy for GO 12329 and 12616. 
We chose the regions that cover at least 3 galaxies (up to 5) per $HST$/WFC3 
pointing. By doing this we were able to observe many galaxies in a small 
number of telescope pointings. We used F125W (hereafter $J_{125}$) 
and $H_{160}$ in these two programs. In GO 12329, the exposure time was one 
orbit per filter per pointing. We found that some galaxies in our sample were 
barely detected with this depth. Therefore in GO 12616, all new WFC3 pointings 
had a depth of two orbits (on-source integration $\sim4800$ sec; see 
Windhorst et al. 2011 for the anticipated sensitivity in two-orbit WFC3 
exposures). We also re-observed most galaxies that were observed in GO 12329, 
so that they have a total of two-orbit depth per band.
As in the NICMOS observations above, the WFC3 observation of each pointing 
(per band) was split into four exposures. Sub-pixel dithering during the 
exposures was also performed to improve the PSF sampling and to remove bad 
pixels and cosmic rays. The transmission curves of the three WFC3 filters
are plotted in Figure 1.

As we mentioned above, several galaxies in our sample were found in the SXDS.
They were covered by the UKIDSS Ultra-Deep Survey (UDS). Their $HST$ WFC3
near-IR data were obtained from the Cosmic Assembly Near-infrared Deep
Extragalactic Legacy Survey \citep[CANDELS;][]{gro11,koe11}.
The exposure depth of the CANDELS UDS data is 1900 sec in 
the $J_{125}$ band and 3300 sec in the $H_{160}$ band. They are shallower
than our WFC3 data for the SDF.

In Appendix A we show the thumbnail images of the galaxies in the two $HST$ 
near-IR bands $J_{110}$ (or $J_{125}$) and $H_{160}$. Note that the 
$J_{125}$-band image of the $z=6.96$ LAE (no. 62) was obtained from GO program
11587 \citep{cai11}. This image also covers another two objects (no. 2 and
no. 59) in our sample. Due to the high sensitivity of WFC3, the majority of
the galaxies in our sample are detected with high significance in the near-IR
images. Only 15 of them --- among the faintest in the optical bands --- have
weak detections ($<5\sigma$) in the $J$ band.

Our WFC3 data processing began with individual flat-fielded, flux calibrated 
WFC3 infrared exposures delivered by the $HST$ archive. Each pointing used a
standard 4-point dither pattern to obtain uniform half-pixel sampling. All 
exposures for each pointing were combined using the software 
{\tt Multidrizzle} \citep{koe02}, with an output plate scale of $0\farcs06$ 
per pixel and a {\tt pixfrac} parameter of 0.8. For our observations, this 
provides Nyquist sampling of the PSF and relatively uniform weighting of 
individual pixels. RMS maps were also derived from the {\tt Multidrizzle} 
weight-map images, following the procedure used in \citet{dic04} that accounts 
for correlated noise in the final images.

Our NICMOS NIC3 data were reduced using a fully automated pipeline NICRED
\citep{mag07}. NICRED starts its processing with raw science data and 
calibration files, and sequentially runs a set of steps from basic calibration 
to final mosaic generation. The most important part of the reduction is the
basic calibration, because the NICMOS data suffer from variable bias, 
electronic ghosts, and cosmic-ray persistence. The basic procedure is as 
follows. NICRED first removes electronic ghosts, handles variable bias, 
rejects cosmic rays, and subtracts sky background. It then removes possible 
cosmic-ray persistence and residual instrument signatures. The rest of 
the procedure are common steps, including flat fielding, a non-linearity 
correction, image alignment and registration, etc. It finally creates 
drizzled images and the corresponding weight maps. The pixel size in the 
final products is $0\farcs1$, a half of the native size.

We list the near-IR photometry of the galaxies in Columns 8 to 10 in Table 1.
All galaxies were observed in the $H_{160}$ band and one or two of the $J$
bands ($J_{110}$ and/or $J_{125}$). 
We perform photometry using {\tt SExtractor}
in dual-image mode with RMS map weighting. The $J$-band is used as the
detection image. A single set of parameters is used for detection in all 
images. We compute the total magnitudes of the objects in both filters by 
using MAG\_AUTO elliptical apertures, with a Kron factor of 1.8 and a 
minimum aperture of 2.5 semi-major radius, then applying aperture corrections. 
The aperture corrections are calculated by running {\tt SExtractor} again with 
five Kron factors between 2.0 and 4.0, analyzing thousands of sources in the 
range $J<26$ mag. We calculate a sequence of median correction vs. Kron 
factor, which we extend to infinite radius with an exponential fit.
All final photometry apertures are visually inspected, to screen against
contamination from nearby objects or spurious detections.
Detections below $5\sigma$ in the $J$ band, or detections below $3\sigma$ in 
the $H$ band, are not listed in Table 1.

\subsection{$Spitzer$ Mid-IR Imaging Data}

We obtained $Spitzer$ IRAC imaging data for the SDF from two GO programs 40026
(PI: E. Egami) and 70094 (PI: L. Jiang). In GO 40026, we observed 20 luminous 
galaxies at $z\ge6$, which were also the targets of our $HST$ program 11149. 
The observations were made in IRAC channel 1 (3.6 $\mu$m) and channel 2 (4.5 
$\mu$m). The other two channels (5.8 and 8.0 $\mu$m) were observed 
simultaneously. We used the cycling dither pattern with medium steps and a 
frame exposure time of 100 sec. The on-source integration time per channel per 
pointing is 3 hr. In GO 70094 (during the $Spitzer$ warm mission), 
we modified our observing strategy and observed 
high galaxy surface density regions, so that we can observe a large number of 
galaxies with slightly more than 10 telescope pointings. The cycling dither 
pattern with small steps was used with a frame exposure time of 100 sec. We 
focused on IRAC channel 1 and increased on-source integration time to $\sim$6 
hr per pointing. The two programs imaged roughly 70\% of the SDF to a depth of 
3--6 hours. They covered all the 62 SDF galaxies in channel 1 and 51 (out of 
62) galaxies in channel 2.

The IRAC Data processing began with the cBCD (corrected BCD) products and 
associated mask and uncertainty images delivered by the $Spitzer$ Science 
Center (SSC) archive. The image fits headers were updated to match the 
astrometry in the SDF optical images. Then the images were processed, 
drizzled, and co-added using using the SSC pipeline {\tt MOPEX}. This is a 
fully automated procedure. The final images have a pixel size of $0\farcs6$, 
roughly a half of the IRAC native pixel scale. The IRAC images for the SXDS 
galaxies were obtained from the SSC archive (programs 60022 and 80057). The 
data were reduced in the same way above. In Appendix (the last two columns) 
we show the thumbnail images of the galaxies in the two IRAC channels.

Mid-IR photometry is complicated by source confusion.
Galaxies in deep IRAC images are often blended with nearby neighbors, so
accurate photometry requires proper deblending and removing neighbors.
We will present IRAC photometry of our galaxies in the second paper of this
series.

\section{REST-FRAME UV CONTINUUM PROPERTIES}

With the narrow- and broad-band photometry from the NB816 to the $H_{160}$
band, we are able to measure the spectral properties, such as \lya\ flux and 
UV continuum slopes, for most of the galaxies in this sample. In order to 
measure the rest-frame UV properties, we produce a model spectrum for each 
of these galaxies. The model spectrum $f_{\rm gal}$ (in units of 
$\rm erg\ s^{-1}\ cm^{-2}\ A^{-1}$, like commonly used $f_\lambda$) is the sum 
of a \lya\ emission line and a power-law UV continuum with a slope $\beta$,
\begin{equation}
	f_{\rm gal} = A \times S_{\rm Ly\alpha} + B \times {\lambda}^{\beta},
\end{equation}
where $A$ and $B$ are scaling factors in units of 
$\rm erg\ s^{-1}\ cm^{-2}\ A^{-1}$, and $S_{\rm Ly\alpha}$ (dimensionless) is 
the \lya\ emission line profile. In the redshift range and wavelength range 
considered here, other nebular lines such as \heii\ can be safely ignored 
\citep[e.g.,][]{cai11}. We first determine the continuum parameters $B$ and 
$\beta$ by fitting the continuum 
$f_{\rm con,\lambda} = B \times {\lambda}^{\beta}$ 
(or $f_{\rm con,\nu} \propto {\lambda}^{\beta+2}$)
with the broad-band data that do not cover the \lya\ emission. Because AB 
magnitude $m_{\rm AB}$ and $\rm Log(\lambda)$ have the simplest linear 
relation $m_{\rm AB} \propto (\beta+2) \times {\rm Log}(\lambda)$, the 
continuum fitting is performed on $m_{\rm AB}$ and $\rm Log(\lambda)$ using 
the {\tt IDL} routine {\tt linfit.pro}. This routine takes $\rm Log(\lambda)$, 
$m_{\rm AB}$, and the errors in $m_{\rm AB}$, and performs a standard 
(weighted) least-squares linear regression by minimizing the chi-square error 
statistic. The errors propagate to the final parameters.

In order to reliably measure continuum slopes, we only consider the galaxies 
with $>5\sigma$ detections in the $J$ band (the deepest band). We further 
require that these galaxies have $>3\sigma$ detections in at least one more 
broad band that does not cover the \lya\ emission. Detections below $3\sigma$ 
are not used here. For the galaxies at $z<6.1$, four broad bands $z'yJH$ can 
be used. As shown in Table 1, the $z<6.1$ galaxies with $>5\sigma$ 
detections in the $J$ band and with $>3\sigma$ detections in another band
are usually detected ($>3\sigma$) in the other two bands, so all four broad 
bands are used for most of the galaxies at $z<6.1$.
For the galaxies at $z>6.1$, the $z'$ band covers \lya, so only three broad 
bands $yJH$ can be used. However, almost half of the $z>6.1$ galaxies with 
$J$-band detections ($>5\sigma$) are not detected at $>3\sigma$ in the 
$y$ band. For these galaxies, we use only the $JH$ bands to derive their UV 
slopes. The information (or upper limits) from the $y$-band photometry is
ignored, because the $JH$-band images are much deeper (by 1--1.5 mag) than
the $y$-band images, and the $y$-band upper limits provide little additional
constraint on $\beta$. 
In addition, in rare cases where some galaxies in the crowded
optical images are significantly affected by their environments like bright 
neighbors (e.g., no. 2 and no. 4), their optical data are not used.

We then determine parameter $A$ for \lya\ emission line. \citet{kas11} 
generated two composite \lya\ emission line profiles for their $z\simeq5.7$ 
and 6.5 LAEs, and found that the two profiles were very similar, showing no 
significant evolution from $z\simeq5.7$ to 6.5 \citep[see also][]{hu10}. 
We use the two composite line profiles as our model \lya\ 
emission lines in Eq. 1: we use the $z\simeq5.7$ profile for the galaxies at 
$z\le6.2$ and the $z\simeq6.5$ profile for the galaxies at $z>6.2$. The shape 
of the \lya\ emission lines has negligible impact on our results, because even
the three narrow-band filters used for \lya\ detection are much wider than the 
typical \lya\ line width. 
We apply IGM absorption to the model spectrum (the IGM absorption has 
already been taken into account in the composite \lya\ emission lines).
The neutral IGM fraction increases dramatically from $z=5.5$ to $z=6.5$, 
causing complete Gunn-Peterson troughs in some $z>6$ quasar spectra 
\citep{fan06}. It is thus important to include IGM absorption in the model 
spectrum. We calculate IGM absorption in the same way as \citet{fan01}.
The IGM-absorbed model spectrum is convolved with the total system response, 
such as filter transmission and CCD quantum efficiency. Finally, the  
\lya\ emission flux (or parameter $A$) is calculated by matching the model
spectrum to the narrow-band photometry (for LAEs) or the $z'$-band photometry 
(for LBGs).
For LBGs, we will see in Section 4.1 that even small $z'$-band photometric 
errors will cause large uncertainties on measurements of $A$, so we adopt the 
\lya\ flux of the LBGs from the galaxies discovery papers, which were 
derived from deep optical spectra.

Figure 2 illustrates our procedure by showing our model fit to the photometric 
data points of two LAEs at $z\simeq5.7$ and $z\simeq6.5$, respectively. For 
the $z\simeq5.7$ LAE in this specific case, we use the photometric data in the 
$z'yJ_{110}H_{160}$ bands (red circles) to fit the model continuum. When the 
continuum (strength $B$ and slope $\beta$) is derived, we apply IGM absorption.
The continuum flux at the blue side of the \lya\ line (the dashed line in the 
figure) is absorbed. Finally the factor $A$ is determined by scaling the \lya\
emission line to match the NB816-band photometry. For the $z\simeq6.5$ LAE in 
Figure 2, we use the data in the $yJ_{125}H_{160}$ bands to compute its
continuum, and use the NB921-band photometry to measure its \lya\ emission.
After $A$, $B$, and $\beta$ are determined, other physical quantities such as 
\lya\ and UV luminosities, rest-frame \lya\ equivalent width (EW), 
and star formation rates (SFRs) are also calculated. The results are shown
in Table 2.

\subsection{UV Continuum Slope $\beta$}

The rest-frame UV-continuum slope provides key information to constrain young 
stellar populations in galaxies. For $z\ge6$ LBGs in the literature, 
especially those selected from $HST$ photometric samples, their UV slopes are 
usually measured from two broad bands $J_{125}$ and $H_{160}$ that do not 
cover \lya\ emission. Recent studies show that high-redshift LBGs generally 
have steep UV slopes and low dust content 
\citep[e.g.,][]{bou09,bou12b,dun12a,fin12,gon12,walter12}. 
For example, \citet{bou09} found that $\beta$ decreases from $\beta\simeq-1.5$ 
at $z\simeq2$ to $\beta\simeq-2$ at $z\simeq6$. \citet{fin12} showed that 
$\beta$ changed from $\beta\simeq-1.8$ at $z\simeq4$ to $\beta\simeq-2.4$ at 
$z\simeq7$. Extremely steep slopes of $\beta\simeq-3$ have been reported 
in very faint $z\simeq7$ galaxies \citep[e.g.,][]{bou10}. 
It was also found that lower-luminosity galaxies tend to have steeper slopes 
\citep[e.g.,][]{bou12b,gon12}. Recently the results on the blue colors in 
high-redshift LBGs have been questioned by \citet{dun12a} and 
\citet{mcl11}, who claimed that previous measurements were significantly 
affected by contaminants in the photo-selected LBG samples and the low 
signal-to-noise (S/N) ratios of faint objects, and that the extremely steep 
slopes were caused by the so-called ``$\beta$ bias''. This bias is partly 
produced by the removal of galaxies with larger (redder) $\beta$ values from 
the sample as lower-redshift interlopers, which would preferentially remove 
flux-boosted sources in the $H_{160}$ band and skew the $\beta$ distribution 
to smaller (i.e., bluer) values as a result. As pointed out by \citet{bou12b}, 
this $\beta$ bias can be large if the same bands ($J_{125}$ and $H_{160}$) are 
also used to select LBGs. With a more restricted selection of LBGs candidates 
from the same $HST$ data set, \citet{dun12a} and \citet{mcl11} found that the 
average slope of $z\simeq7$ LBGs is $\beta\simeq-2$, which is not steeper than 
the bluest galaxies at $z\le4$. They further found no relation between $\beta$ 
and the UV continuum luminosity. As the debate is still going on 
\citep[e.g.,][]{bou12b,fin12,dun12b}, the key is to construct a high-redshift 
galaxy sample that is free from such a bias.

Our galaxy sample presented here is unlikely to be strongly 
affected by this $\beta$ 
bias, not only because these galaxies have spectroscopic redshifts but also 
because our original source selection did not utilize any information on the 
rest-frame UV continuum slope. Furthermore, we did not include weak detections 
in our analysis. For many galaxies, especially the $z\simeq5.7$ LAEs and 
$z\simeq6$ LBGs, $\beta$ was measured from more than two bands. 
It should be pointed out, however, our LBGs were spec-confirmed to have strong 
\lya\ emission, which could bias the LBG sample to galaxies with lower 
extinction and thus to those with bluer UV continuum slopes.

Figure 3(a) shows the rest-frame UV continuum slopes $\beta$ as a function of 
$M_{1500}$, the absolute AB magnitude of the continuum at rest-frame 1500 \AA. 
$M_{1500}$ is calculated from $B\times{\lambda}^{\beta}$ in Eq. 1. Figure 3
only includes the galaxies that have measured $\beta$ in Table 2. These 
galaxies cover the brightest UV luminosity range $M_{1500}<-19.5$ mag.
For the galaxies that do not have measured $\beta$, they do not have reliable
detections ($<5\sigma$) in the near-IR bands, so they are usually fainter than 
$M_{1500}=-19.5$ mag (see Section 4.2 and Figure 6). The green circles in
Figure 3(a) represent the LBGs at $z\simeq6$, and the blue and red circles 
represent the LAEs at $z\simeq5.7$ and 6.5 (including $z\simeq7$), 
respectively. The values of $M_{1500}$ and $\beta$ are also listed in Columns
2 and 3 of Table 2. Our measurements of $\beta$ are reliable because of the 
secure redshifts and multi-band photometry. As we addressed above, the slopes 
are calculated from all available broad-band data that do not cover the \lya\ 
emission line. For any $\beta$ that was derived from more than two bands, its 
error is usually smaller than 0.4. So most of the $z\simeq5.7$ LAEs and 
$z\simeq6$ LBGs have relatively small errors on $\beta$. For the $z\simeq6.5$ 
LAEs, their slope errors are generally larger, as many of them were only 
detected in the two near-IR bands in addition to NB921.

The galaxies in Figure 3 apparently have steep UV-continuum slopes that are
roughly between $\beta\simeq-1.5$ and $\beta\simeq-3.5$. The weighted mean is
$\beta=-2.29$, and the median value is $\beta=-2.35$. 
This slope is slightly steeper than those from previous 
observations \citep[e.g.,][]{bou12b,dun12a,dun12b,fin12} and 
simulations \citep[e.g.,][]{finlator11,day12} in the literature.
The slope $\beta$ also appears to show weak dependence on $M_{1500}$ that 
lower-luminosity galaxies have steeper slopes. The dashed line is the best 
linear fit to all the data points. This trend is caused by the several most 
luminous galaxies at $M_{1500}\simeq-22$ mag. These luminous galaxies have 
red colors $\beta\simeq-1.8$. If we exclude these galaxies in our fit, the 
trend almost disappears. The dash-dotted line in the figure illustrate the 
result by displaying the best fit to the galaxies at $M_{1500}>-21.7$ mag,
and its slope is consistent with zero.

In the above analysis, galaxies with near-IR detections below $5\sigma$ in the 
$J$ band 
were discarded. 
This has negligible impact on our results. Our $HST$ near-IR data have 
relatively uniform depth (two orbits per band per pointing), so the $5\sigma$ 
detection cut indeed puts a flux limit on $M_{1500}$ in Figure 3(a), which 
does not affect our results in the bright region $M_{1500}<-19.5$ mag. We also 
perform two simple tests to investigate whether large errors on photometry or 
$\beta$ have strong impact on our results. In the first test we stack the 
images of 9 galaxies whose errors of $\beta$ ($\sigma_\beta$) are greater than 
0.8 in Figure 3, and then measure photometry and calculate $\beta$ 
on the stacked images in the $J_{125}$ and $H_{160}$ bands. The input images 
are scaled so that the galaxies have the same magnitude in the $J_{125}$ band 
before stacking, otherwise the final image would be dominated by the brightest 
objects. The stacked images have much higher signal-to-noise (S/N) ratios. 
The 9 galaxies have a median slope of $\beta\simeq-2.1$. The slope for the 
stacked object is $\beta=-2.2\pm0.2$, which agrees with the median value of 
the individual galaxies. The result is shown as the magenta star in Figure 4, 
where the gray circles represent the same galaxies shown in Figure 3(a).
In the second test there are 10
galaxies with $\beta<-2.8$ that have both $J_{125}$ and $H_{160}$ images.
The median value of their slopes is $\beta\simeq-3.0$.
We stack their images in the same way as we did above. The slope for the 
combined object is $\beta=-2.9\pm0.2$, which is also consistent with the 
median value of the individual galaxies. The magenta square in Figure 4 
represents the combined object. From these tests, our measurements of $\beta$
are not biased even for faint objects with relatively large photometric errors.

\subsection{$\beta$-$M_{1500}$ Relation}

It is generally believed that high-redshift galaxies have steep UV slopes
$\beta$, and $\beta$ changes with redshift: higher-redshift galaxies tend to
be bluer due to less dust extinction. The relation between $\beta$ and UV
luminosity ($M_{\rm UV}$, or $M_{1500}$ in this paper) is still controversial.
In Figure 4 we compare our results with recent studies
\citep{bou12b,dun12a,fin12}.
\citet{bou12b} found a correlation between $\beta$ and
$M_{\rm UV}$ at all redshifts in a range of $4<z<7$. They claim that galaxies
have bluer UV colors at lower luminosities. The green triangles in Figure 4
represent the average slopes at $z\sim6$ from their study.
\citet{dun12a} used a more restricted selection of LBGs by excluding low S/N
detections from a similar $HST$ data set. They found that high-redshift
galaxies have an average slope of $\beta\simeq-2$, and it does not change with
$M_{\rm UV}$. The blue crosses in Figure 4 represent the average slopes
at $5<z<7$ from their study. \citet{fin12} also found that $\beta$ shows minor
relation with $M_{1500}$ in the redshift range from $z=4$ to 8. The red
diamonds in Figure 4 represent the average slopes at $z\simeq6$ from their
study.

Figures 3 and 4 suggest that in our sample there is little dependence of
$\beta$ on $M_{1500}$ in the range of $M_{1500}<-19.5$ mag. The three studies
mentioned above were all based on $HST$ data that are $\sim1-2$ mag deeper
than ours. In the luminosity range of $M_{1500}<-19.5$ mag, these studies do
not show evidence of a significant correlation either, though there is likely
a weak trend at the fainter range $M_{1500}>-19.5$ mag. Therefore, it is
possible that the reported correlation between $\beta$ and $M_{\rm UV}$,
if exists, is mainly driven by very faint galaxies.

The mean value of our measured $\beta$ is --2.3. It is slightly steeper than
$\beta=2.1-2.2$ by \citet{bou12b} in the same luminosity range, and is also
steeper than $\beta\simeq-2$ by \citet{dun12a} and \citet{fin12}. This is due
to the nature of our sample. The galaxies in the previous studies are
photo-selected LBGs. Our galaxies are spectroscopically confirmed and have
strong \lya\ emission lines, suggesting that they likely have lower dust 
extinction, and thus steeper UV slopes.

\subsection{Slopes in LAEs}

It is not clear whether high-redshift LAEs and LBGs represent physically
different galaxy populations. Early observations found that LAEs at $z=5\sim6$
are smaller and less massive compared to LBGs, and thus constitute younger
stellar populations \citep[e.g.,][]{dow07,pir07}. However, this apparent
difference could be a direct result of selection effects, and LAEs may
represent a subset of LBGs with strong \lya\ emission lines
\citep[e.g.,][]{day12}.
Whether or not LAEs and LBGs have similar populations will be reflected in
their physical properties. For example, if LAEs are substantially younger,
they are expected to have steeper UV slopes.

In Figures 3 and 4 the distribution of our slopes $\beta$ for the $z\simeq5.7$
and $z\simeq6.5$ LAEs are $\beta=-2.3\pm0.5$ and $\beta=-2.0\pm0.8$,
respectively. The $\beta$ distribution for the LBGs is $\beta=-2.5\pm0.4$.
Although the scatters are large, the LAE slopes are apparently not steeper
than the LBG slopes in our sample. They are also not steeper than the slopes
of the LBGs in previous studies mentioned above, which is contrary to the
assumption that LAEs contain younger stellar populations than LBGs do.
This indicates that most LAEs are probably not as young as what was previously
suggested. 

\subsection{Galaxies with Extreme UV Slopes}

The majority of our galaxies have slopes $\beta>-2.7$ in Figure 3(a). There 
are a fraction of galaxies whose slopes are steeper than --2.7. Some of them
have relatively small errors around $\sigma_\beta\sim0.4$. The existence of 
these $\beta\simeq-3$ galaxies are statistically robust (not a distribution 
tail). In Figure 3(b) our statistical test shows that the distribution of the 
observed $\beta$ is broader and flatter than the distribution of $\beta$ 
expected if all the galaxies have an intrinsic $\beta=-2.3$ with the observed 
uncertainties. It clearly indicates a statistically significant excess of
galaxies with $\beta\simeq-3$. These galaxies
do not show distinct properties in many aspects such as \lya\ EW and UV
morphology. The existence of these $\beta\simeq-3$ galaxies in our sample is
intriguing. The UV colors cannot be arbitrarily blue for a given
initial mass function (IMF). For the commonly used Salpeter IMF, a
$\beta\simeq-3$ usually means a very young stellar population with extremely
low metallicity and no dust content, regardless of star-formation histories
\citep[e.g.,][]{finlator11,wil11}. In particular, the $\beta<-3$ galaxies,
if confirmed by future deeper observations, will be challenging for current
simulations \citep{finlator11}. In addition, the broad distribution of the 
observed $\beta$ also suggests significant intrinsic scatter in dust and/or 
age. Detailed discussion is deferred to
the next paper when we perform SED modeling with the IRAC data.

From Figure 3 the five most luminous galaxies at $M_{1500}\simeq-22$ mag
have UV slopes $\beta\simeq-1.8$, which is significantly redder than the
mean value $\beta=-2.3$ of the sample. The measurements of $\beta$
are robust since these galaxies are very bright and detected in
almost all the broad bands redward of \lya. In particular, they are all
detected at high significance in the IRAC bands.
Their \lya\ emission is relatively small compared
to their UV continuum emission. From the $HST$ images, they have very extended
morphologies, even with double or multiple cores, suggesting
that they are interacting systems with
enough amounts of dust to redden their UV colors.
These galaxies are similar to the luminous galaxies at $z\sim6$ found by
\citet{wil13} in the CFHT Legacy Survey. \citet{wil13} selected a sample of
luminous galaxies with $M_{\rm UV}\simeq -22$ in a 4 sq. degree area and found
that their galaxies have very red colors with a typical slope $\beta=-1.4$,
which is caused by large dust reddening of $A_V>0.5$.

\section{\lya\ PROPERTIES}

\subsection{\lya\ Luminosity}

Spectroscopic data generally provide reliable line luminosities, but this
becomes complicated for $z\ge6$ galaxies. During the
spectroscopic observations of these galaxies, the total integration time is at
least a few hours per object or per slit mask. Sometimes the images of one
object are taken in different nights. Light loss varies with many parameters,
such as varying seeing, different observing conditions, possible offsets
between targets and slits, and even the sizes of targets. So it is 
difficult to accurately correct for light loss.
In addition, even with a few hour integration on the largest telescopes,
the spectral quality of many galaxies are still poor, simply because they are
faint. All of the above effects can easily result in large uncertainties in 
the measurements of their \lya\ luminosities.

Photometric data provide an alternative way to measure the observed \lya\ 
flux, especially for LAEs. We have high-quality narrow-band and broad-band 
photometry and secure redshifts. Our 
procedure used all available data and was straightforward. The \lya\ 
luminosities of the LAEs were derived from narrow-band photometry after their 
underlying continua were subtracted. For the majority of the LAEs in our 
sample ($\rm EW\ge20$\AA), their narrow-band photometry is dominated by \lya\ 
emission, meaning that the real \lya\ emission is smaller --- but not much 
smaller --- than the total narrow-band photometry. Therefore, we neither
significantly overestimate the \lya\ luminosity (the maximum is the total
narrow-band emission), nor significantly underestimate it (the minimum is at
least a half of the maximum for most LAEs).

The measurements of our \lya\ luminosities for LBGs are less accurate.
For most of the LAEs, their narrow-band photometry is dominated by the \lya\
emission, so their \lya\ luminosities can be well determined. For the LBGs,
\lya\ luminosities were calculated from the $z'$-band photometry. The $z'$
band is very broad, and its photometry is almost completely dominated by the
UV continuum flux. 
Therefore, even small uncertainties on the $z'$-band photometry or on
measurements of UV continuum will cause very large uncertainties on the \lya\
emission. Therefore for the \lya\ luminosities of our LBGs, we directly
adopt the values from the galaxy discovery papers, which were derived from
optical spectra. All the measured \lya\ luminosities are given in Column 4 of
Table 2.

Figure 5 compares our \lya\ luminosities of the LAEs to those given in 
\citet{kas11}. The upper panel shows the comparison between our luminosities
and the luminosities from the photometric data in the galaxy discovery 
papers. The photometric data used in these papers are usually the narrow-band 
and $z'$-band photometry. Our results are well consistent with these previous
measurements with a small scatter. The lower panel shows the comparison 
between our measurements and the measurements from the spectroscopic data in
\citet{kas11}. The two results also agree with each other, although our 
measurements are slightly larger and the overall scatter is larger than that 
in the upper panel.

\subsection{\lya\ EW}

Column 5 of Table 2 lists the \lya\ EWs of the galaxies that have measured UV 
continuum properties in Section 3.1. These galaxies are also shown as filled
circles in Figure 6. The bluest galaxy with an unusual $\beta=-4.37$ is 
excluded. We also estimate EWs for the LAEs that 
have weak or none detections ($<5\sigma$) in the $J$ band. These galaxies 
could potentially have very large EWs. In order to estimate their EWs (or the 
lower limits of EWs), we assume a UV slope $\beta=-2.3$ and compute UV 
flux from their $J$-band photometry. Almost all these galaxies are detected
at $>2\sigma$ in the $J$ band. If a detection is below $2\sigma$, we use 
$2\sigma$ as an approximation. The measurements are shown as the open circles
in Figure 6. 
The upper panel shows the relation between the \lya\ luminosity and UV
continuum luminosity $M_{1500}$. The \lya\ luminosity has a weak dependence on 
$M_{1500}$. The lower panel displays the measured \lya\ EWs as a function of
$M_{1500}$. Our galaxies have moderately strong EWs in the range from $\sim$10 
to $\sim$200 \AA. The median EW value of LAEs is 81 \AA, consistent with those 
given in \citet{kas11}. As expected, the LAEs with weak near-IR 
detections have much higher EWs roughly between 100 and 300 \AA. 

Figure 6 (the lower panel) shows that lower-luminosity galaxies tend to have 
higher EWs. It has been a long-standing debate whether this is a real 
relation, or just the result of a selection effect, i.e., in a flux-limited 
survey, lower-luminosity galaxies naturally have larger EWs in order to be 
detected in imaging data or identified in spectroscopic data. Many papers have 
reported this inverse relation between \lya\ EW and UV luminosity to be real 
at $z\ge3$ \citep[e.g.,][]{sha03,and06,red08,sta10}, while others claimed that 
the relation is not obvious, when the selection effects are properly taken 
into account \citep[e.g.,][]{nil09,cia12}. In Figure 6 there is an apparent 
lack of large-EW LAEs at the high-luminosity end that does not suffer from 
this selection effect \citep{and06}, suggesting that the relation in our 
sample is likely real. However, the relation should be weaker than it appears, 
because we have missed faint galaxies with small \lya\ EWs by selection. 
We will further assess the effects of the source selection on the shape
of the EW(\lya)--$M_{\rm UV}$ relation in Section 5.1.
As we will mention in the next subsection, one explanation for the relation
between \lya\ EW and UV luminosity (if exists) is that the escape fractions of 
\lya\ photons in higher-SFR (i.e., higher UV luminosity) 
galaxies are smaller \citep[e.g.,][]{for12,gar12}.

\subsection{UV and \lya\ SFRs}

We estimate SFRs from the UV continuum luminosity and \lya\ luminosity, 
respectively. The UV SFR is derived by:
\begin{equation}
   \rm SFR (UV) = 1.4 \times 10^{-28}\ L_{\nu}(UV)\ M_{\sun}\ yr^{-1},
\end{equation}
where L$_{\nu}$(UV) is the UV continuum luminosity \citep{ken98,mad98}. We use
the average of the luminosities at rest-frame 1500 and 3000 \AA\ to compute
L$_{\nu}$(UV). For the \lya\ SFR, we use the expression:
\begin{equation}
   \rm SFR (Ly\alpha) = 9.1 \times 10^{-43}\ L(Ly\alpha)\ M_{\sun}\ yr^{-1},
\end{equation}
which is based on the line emission ratio of \lya\ to \ha\ in Case B 
recombination \citep{ost89} and the relation between SFR and the \ha\ 
luminosity \citep{ken98}. The derived SFRs are listed in Columns 6 and 7 of 
Table 2.

Figure 7 compares the two estimated SFRs without correction for dust 
extinction. As in Figure 6, the filled circles are the galaxies with reliable 
measurements. These galaxies have moderate SFRs from a few to a few tens 
$M_{\sun}$ yr$^{-1}$. The open circles are the galaxies with weak near-IR 
detections, and their measurements of SFRs are crude. Because they have large 
\lya\ EWs and weak UV continuum flux (Figure 6), they show small UV SFRs 
($\sim1-2$ $M_{\sun}$ yr$^{-1}$) and high SFR(\lya)-to-SFR(UV) ratios.
For the purpose of comparison, we correct SFR(\lya) for the IGM absorption.
\lya\ emission at high redshift is complicated by factors such as the resonant
scattering of \lya\ photons and the IGM absorption, and it is difficult to 
model \lya\ radiative transfer and predict intrinsic \lya\ emission
\citep[e.g.,][]{zheng10,dijkstra11,yajima12,silva13}.
We thus do not correct individual galaxies. Instead, we derive an average
\lya\ flux reduction for all the galaxies. The average correction is 
calculated from the composite \lya\ line of \citet{kas11} by assuming that the 
intrinsic \lya\ profile is symmetric around the peak of the composite line.
The dotted line in Figure 7 illustrates the average correction of 24\%.
When the effect of the IGM absorption is taken into account in this way, the 
UV and \lya\ SFRs for most luminous galaxies (filled circles) are reasonably 
consistent within a factor of two.

Figure 7 shows a trend that the SFR(\lya)-to-SFR(UV) ratio becomes 
increasingly smaller towards higher SFRs. The dashed line illustrates this 
trend by displaying the best log-linear fit to the data points. This trend is 
already suggested by the relation between EW(\lya) and $M_{1500}$ in Figure 6, 
since SFR(UV) reflects $M_{1500}$ and EW(\lya) mimics SFR(\lya)/SFR(UV).
Accordingly, this trend has also been shaped by the selection effect due to
the nature of the flux-limited sample, i.e., we have missed low SFR(UV) 
galaxies with low SFR(\lya)-to-SFR(UV) ratios. Therefore, the real trend of 
the SFR(\lya)-to-SFR(UV) ratio, if exists, should be much weaker than it 
appears. This trend could also be due partly to the relation between SFRs and
the escape fractions of \lya\ photons. The model of \citet{gar12} predicts 
that at high redshift, the escape fractions of \lya\ photons in low-SFR 
galaxies are close to 1, but become smaller in galaxies with higher SFRs
\citep[see also e.g.,][]{for12}. This is also consistent with \citet{cur12}, 
who found that in a sample of luminous (high SFRs) $z\ge6$ LBGs, the \lya\ 
SFRs are only $\sim40$\% of the UV SFRs.

\section{DISCUSSION}

\subsection{\boldmath EW(\lya) -- $M_{1500}$ Correlation for LAEs}

Figure~6 shows an apparent correlation between EW(\lya) and $M_{1500}$ in our
sample. As we already mentioned, the relation is affected by the nature of the 
flux-limited sample, i.e., the limiting Ly$\alpha$ line flux (and therefore 
the limiting line luminosity) associated with both narrow-band and 
spectroscopic observations. In Figure~8(a), we plot the current LAE samples at 
$z\simeq5.7$ and 6.5, as well as those from another large LAE survey at 
$z\simeq3.1$, 3.7, and 5.7 by \citet{ouc08}. The diagonal dotted line is 
defined by a \lya\ luminosity of $2.5 \times 10^{42}$ erg s$^{-1}$, which 
roughly corresponds to the limits of these surveys. 
The figure illustrates that the slope of the EW(\lya)--$M_{\rm UV}$ relation 
is largely shaped by the limiting luminosity, which censors sources that would 
fall below the diagonal dotted line.
In addition, as pointed out by \citet{cia12}, EW is a derived quantity from
the emission line and underlying continuum, so any relation between EW
and the line (or continuum) flux may suffer from correlated errors.
Such correlated errors could result in an apparent relation by scattering
objects towards one direction \citep{cia12}.

Although nothing definite can be said about the properties of
high-redshift LAEs below our detection limit, it is natural to expect
that the distribution of LAEs extends toward smaller
EW(\lya). Therefore, it would be interesting to see how the population
of low-redshift low-luminosity LAEs compares with the high-redshift
population shown in Figure~8(a). In Figure~8(b), we include the sample of
GALEX-selected $z\simeq0.2-0.4$ LAEs presented by
\citet{cowie10,cowie11b}. The figure shows that this low-redshift LAE
sample contains the kind of LAEs that would be missed at high redshift
(i.e., below the dotted line). Once these low-luminosity LAEs are
included, the EW(\lya)--$M_{\rm UV}$ correlation could diminish. This
is consistent with the fact that no evolution has been established
with the EW distribution of LAEs from $z\simeq0.2-0.4$ to $z\simeq3$
\citep{cowie10} and from $z\simeq3.1$ to 5.7 \citep{ouc08}. This may
suggest the possibility that although bright LAEs are more abundant at
high redshift, the underlying \lya\ EW distribution may be fairly
invariant. However, we also caution that the lack of a correlation in
Figure~8(b) is partly caused by another selection effect in these LAE
samples, namely EW(\lya) $\ge$ 20 \AA, which produces a sharp
horizontal boundary at the bottom. In addition, the mix of different samples 
in Figure 8 could dilute intrinsic relations (if exist).

Given these selection effects, it is difficult to determine with the current 
LAE samples how strong the correlation between EW(\lya) and $M_{\rm UV}$ is.
To perform such an analysis in a
statistically meaningful manner, we will need a much larger
LAE sample, as pointed out by \citet{nil09}. For the discussion in
the subsequent sections, we will assume two extreme cases: (1) the
slope of the EW(\lya)--$M_{\rm UV}$ correlation is as steep as seen in
Figure~6 (the dashed line in the lower panel); (2) the slope of the
EW(\lya)--$M_{\rm UV}$ correlation is completely flat. As we shall
see, these two limiting cases will minimize/maximize the size of the
LAE population and therefore their contribution to the rest-frame UV
luminosity density.

\subsection{LAEs and LBGs}

In this paper we call galaxies found by the narrow-band technique LAEs and 
those found by the dropout technique LBGs. This is a widely used definition.
Strictly speaking, this LAE/LBG classification only reflects the methodology
that we employ to select galaxies. It does not necessarily mean that the two 
types of galaxies are intrinsically different. A galaxy with strong \lya\ 
emission could be detected by the both techniques. 
Another popular definition of LAEs is based on the rest-frame EW of the \lya\ 
emission line. One galaxy is a LAE if its \lya\ EW is greater than, for 
example, 20 \AA. With this definition, almost all the galaxies in our sample 
are LAEs, as seen from Figure 6. This definition with \lya\ EW is physically
more meaningful. However, the measurements of \lya\ EWs are usually
accompanied with large errors. Furthermore, one can easily obtain a
flux-limited sample, but not a EW-limited sample.

It is not yet totally clear whether high-redshift LAEs and LBGs
represent physically different populations. 
Direct comparison between LAEs and LBGs is difficult. The procedure of
obtaining a spectroscopic sample of LAEs is relatively
straightforward. LAE candidates are selected based on their
\lya\ luminosities (and one or more broad-band photometry), and
follow-up spectroscopic identification is also based on their
\lya\ luminosities. This results in a complete flux-limited sample in
terms of \lya\ luminosity. For LBGs at $z\ge6$, candidate selection is
based on broad-band colors, but spectroscopic identification is based
on \lya\ luminosity. Therefore, the resultant LBG sample is
inhomogeneous in depth of either \lya\ luminosity or UV continuum
luminosity, and represents only a subset of LBGs with strong \lya\ emission.

So far, we did not find significant differences
between our LAEs and LBGs in the UV luminosity range of $M_{1500}<-19.5$ mag.
Figures 3 and 4 have shown that they have similar UV continuum slopes. The
LAEs do not exhibit steeper slopes than the LBGs, indicating that
their underlying stellar populations are not very different.
To examine the relation between LAEs and LBGs further, one useful tool
is the EW(\lya)--$M_{\rm UV}$ plot in the form of Figures 6 and 8. We
expect the two populations to have a large overlap, but the focus here
is on the non-overlapping population(s) that can be picked up by one
selection method but not by the other. For example, the LAE selection
may pick up galaxies with extremely large EW(\lya) whose continuum
emission is so faint that they will drop out of LBG samples.
Alternatively, the LBG selection can pick up galaxies with small
EW(\lya) which do not show up in LAE samples. Therefore, we would like
to know how significant/insignificant such non-overlapping populations
between LAEs and LBGs are.

Figure~9 plots EW(\lya) and $M_{\rm UV}$ of high-redshift LAEs (from
Figure~8) and LBGs together. The latter sample comes from this work
($z\simeq6$) and from \citet{sta10} ($z\simeq3-6$). The figure suggests
that there is no significant difference between the LAE and LBG
populations in terms of the EW(\lya) and $M_{\rm UV}$ distributions.
The two populations occupy roughly the same region on the plot. For
example, we do not see any sign of enhanced \lya\ strengths among
the LAE population. This implies that the LBG selection would almost
fully recover the LAE population. In other words, there are very few
LAEs with exceptionally large \lya\ EWs that would drop out of
the LBG selection due to their faintness in continuum.
The reverse (i.e., LBGs that would escape the
LAE selection due to weak \lya\ emission) is difficult to assess with
our current sample because our spectroscopic program was not designed
as an extensive follow-up of LBGs. In the next section, we will
construct the UV continuum LF of LAEs and examine how it
compares with the UV continuum LF of LBGs.

\subsection{UV Continuum Luminosity Function of LAEs}

Because of the deep $HST$ images, we were able to detect almost all the LAEs 
at a significance of $>2\sigma$. This allows us to derive the UV continuum LF 
of LAEs directly from the data. The UV LF of LAEs together with the UV LF of 
LBGs will help us estimate the fraction of galaxies that have strong \lya\
emission, and constrain the contribution of LAEs to the total UV ionizing 
photons. 
Our LAEs are from a \lya\ flux-limited (not EW-limited) sample, so they have 
different detection limits of \lya\ EWs at different UV luminosities (see
Figures~8 and 9). In this subsection, we will derive the UV LF of LAEs with 
\lya\ EWs greater than 20 \AA, by extrapolating the observed LAE population 
down to this EW threshold.

We first compute the number densities of the LAEs without extrapolating the
LAE population to the EW threshold of 20 \AA. 
The area covered by our LAEs is calculated 
from the area of the whole LAE sample given in \citet{kas11}, by matching the 
number of our LAEs to the number of the whole LAE sample. We do not use the 
actual area that our $HST$ observations covered, because we selected high 
surface-density regions of galaxies for the $HST$ programs (see Section 2.3). 
We then incorporate the completeness corrections from \citet{kas11}. 
Completeness is calculated for individual galaxies, as it is a function of 
narrow-band magnitude. The resultant number densities are shown as the 
thick blue and red lines in Figure 10. They set the absolute lower limits 
on the UV LF of LAEs at $z=5.7$ and 6.5.

We then extrapolate the densities to $\rm EW=20$ \AA\ using the \lya\ EW 
distribution function. Figure 11 shows the \lya\ EW distribution of the LAEs 
in our sample. We assume that the EW distribution is an exponential function 
$n\propto e^{-w/w_0}$, where $w$ is EW and $w_0$ is the $e$-folding width. 
It is often assumed that $w_0$ is correlated with UV luminosity
\citep[e.g.][]{kas11,sta11}. We assume that $w_0$ is a log-linear function of 
$M_{1500}$: ${\rm log}(w_0) = a + b \times (M_{1500}+20)$, where $a$ and $b$ 
are scaling factors. Given the small numbers of the galaxies, it is not 
realistic to obtain reliable measurements for both $a$ and $b$. Therefore, we
consider the two extreme cases mentioned in Section 5.1: (1) $b=0.225$, 
where we assume that the slope of the EW(\lya)--$M_{1500}$ correlation 
is as steep as seen in Figure~6 (the dashed line in the lower panel); 
(2) $b=0$, where the slope of the correlation is completely flat.
These two limiting cases will minimize ($b=0.225$)/maximize ($b=0$)
the size of the LAE population. The real $b$ value is between
0 and 0.225. We then determine $a$ by fitting the exponential
function to the EW distribution of the LAEs with $M_{1500}<-19.5$ mag
in our sample. The best fit is shown in the top panel of Figure 11.
With the derived $a$ and assumed
$b$ we can calculate the \lya\ EW distribution at any luminosity. The
other panels in Figure 11 show the predicted EW distributions for the
two extreme cases with the dashed ($b=0.225$) and dash-dotted ($b=0$) lines.

The shaded regions in Figure~10 show the UV LFs of LAEs corrected to 
$\rm EW(Ly\alpha) = 20$ \AA\ for the two extreme cases. Their lower and upper
boundaries represent the lower and upper limits of the LFs. The $1\sigma$
statistical uncertainties have been included. At $M_{1500}<-20.5$ mag,
no correction is actually applied, because our sample is complete at 
$\rm EW = 20$ \AA\ down to this magnitude. Towards fainter magnitudes, the 
correction becomes increasingly larger, and the difference of correction 
between the two cases also becomes larger. In the faintest end, the difference 
of the number densities between the lower and upper limits is larger than 
a factor of 3.

The cyan and green curves in Figure~10 show the UV LFs of 
photo-selected LBGs from the literature \citep{bou07,bou11,mcl09,ouc09}, 
compared to our results. At the bright end, the LAE UV LFs are roughly 
comparable to the LBG LFs\footnote{Strictly speaking, our first $HST$ GO 
program 11149 slightly favored galaxies with bright continuum flux. This could 
increase the number densities by up to 50\% in the bright end, but is not 
large enough to change the basic conclusion.},
consistent with similar results reported by \citet{ouc08} and \citet{kas11}.
This suggests a large fraction of LAEs among the brightest LBGs. Such high 
fraction has been observed by \citet{cur12}, who found that the LAE fraction 
in a sample of very bright ($M_{\rm UV}\le-21$ mag) UDS LBGs is about 50\%. 
\citet{sta11} reported a lower fraction ($20\pm8$\%) of LAEs with EW(\lya) 
$>25$ \AA\ in a sample of bright $z\sim6$ LBGs ($-21.75<M_{\rm UV}<-20.25$). 
We note, however, 
their sample contains a small fraction (10\%) of LAEs with $M_{\rm UV}<-21$ 
mag, so their LAE fraction is dominated by the galaxies at $M_{\rm UV}>-21$ 
mag. Our LAE fraction ($\sim35$\%) in a similar range of $-21<M_{\rm UV}<-20$ 
is not much different from theirs, given their slightly higher EW limit.
Note that the measurements of the LAE fraction among the brightest 
galaxies are subject to large uncertainties owing to the small numbers of 
galaxies available.

In Figure 10, the UV LF of LAEs at the faint end is even more uncertain, as 
it depends critically on how we extend the EW(Ly$\alpha$) distribution toward
EW = 20 \AA, which is far below our detection limits. Current ground-based
spectroscopy is not yet able to reach this regime. 
Depending on the value of $b$ we assume, the faint-end slope of the LAE UV LF 
could be almost as steep as that of the LBG UV LF ($b\sim0$) or significantly 
flatter ($b>0$). If we maximize the LAE UV LF by assuming $b\sim0$, we have to 
apply a large incompleteness correction ($>\times10$ at the faintest end) to
account for LAEs down to EW(Ly$\alpha$) = 20 \AA. It means
that the \lya\ EW distribution must be highly concentrated on small
EWs, so that the observed LAEs represent a small sub-group of high-EW
galaxies that constitutes a tiny fraction of the overall LAE population.
Despite this uncertainty, it is reasonable to conclude with Figure~10 that 
the number density of LAEs falls significantly short of that of LBGs at the 
faint end. This result could have an important implication for the
population of galaxies that were responsible for cosmic reionization.
It implies the existence of a large population of LBGs with weak \lya\ 
emission ($\rm EW<20$ \AA). Because the UV LF
slope of LBGs is steep, the vast majority of the UV photons would
come from very faint galaxies (far below $L^{\ast}$). This suggests
that the LBG population with weak \lya\ emission dominate the
UV photon budget for cosmic reionization.

In Figure 10, the comparison between the UV LFs of LAEs at $z\simeq5.7$ and 
6.5 is straightforward, and it shows little evolution. 
This supports the previous claim that the \lya\ LF evolution 
from $z\sim5.7$ to 6.5 is due to an increasing
neutral fraction of the IGM between the two redshifts and not due to the
evolution of the LAE galaxy population \citep{kas06,kas11}. These
studies showed that the \lya\ LF evolves rapidly from $z\sim5.7$ to
6.5, and pointed out in particular that there is a lack of luminous
LAEs at $z\sim6.5$. In contrast, the evolution of the UV LF of
LAEs between the two redshifts is much more modest. This was explained
by the increasing neutral fraction of IGM from $z\sim5.7$ to 6.5 that
attenuates \lya\ emission. In these studies, the UV continuum
luminosities were measured from the SDF $z'$-band imaging data, which
can only detect bright LAEs, and are also contaminated by \lya\ emission 
(and Lyman break) for $z\simeq6.5$ LAEs. Our $HST$ data were deep enough
to detect almost all the LAEs in our sample, so our measurements of UV
continuum luminosities are more robust. We confirm the consistency of
the UV LFs of LAEs at $z\simeq5.7$ and 6.5, which rules out the
possibility of a strong intrinsic galaxy evolution between the two
redshifts, and strengthens the interpretation that the increasing
neutral fraction of IGM causes the strong evolution of the \lya\ LF
towards $z\simeq6.5$.

\section{SUMMARY}

We have carried out deep $HST$ near-IR and $Spitzer$ mid-IR observations of a 
large sample of spec-confirmed galaxies at $5.6\le z\le7$. The sample contains 
51 LAEs and 16 LBGs, representing the most luminous galaxies in this redshift 
range. The LAEs were from a complete flux-limited sample. The LBGs have
quite different depth, and only contain those with strong \lya\ emission.
The majority of the galaxies (62 out of 67) were discovered in the SDF, and 
the remaining were found in the SXDS. The $HST$ observations of the SDF 
galaxies were made with WFC3 and NICMOS in the $J$ (F110W or F125W) and $H$ 
(F160W) bands. The depth is two $HST$ orbits per band for most of the objects. 
With such depth nearly 80\% of the galaxies were detected at high significance 
($>5\sigma$) in the $J$ band. The $Spitzer$ observations of the SDF were made 
in two IRAC channels 1 and 2. The depth varies across the field from $\sim$3 
hrs to $\sim$6 hrs. The infrared data of the five SXDS galaxies were taken from 
the $HST$ and $Spitzer$ archive. In addition to the infrared data, 
we also have extremely deep optical images in a series of broad bands 
($BVRi'z'y$) and narrow bands (NB816, NB921, and NB973). We have used the 
combination of the optical and infrared data to derive the properties of 
rest-frame UV continuum and \lya\ emission, such as UV continuum luminosities 
and slopes, \lya\ luminosities and EWs, and SFRs etc. 

While the whole sample covers a large UV continuum luminosity range from 
$M_{1500}\le-22$ mag to $M_{1500}\ge-18$ mag, the galaxies with significant
detections ($>5\sigma$) in the $J$ band cover a bright range of
$M_{1500}\le-19.5$ mag. These galaxies have steep UV continuum slopes 
roughly between $\beta\simeq-1.5$ and $\beta\simeq-3.5$, with a weighted mean
of $\beta\simeq-2.3$. This value is slightly steeper than the slopes of 
photo-selected LBGs reported in previous studies in the literature. This is 
due to the fact that our galaxies have strong \lya\ emission, and thus lower 
dust extinction and steeper UV slopes. The slope $\beta$ shows little trend 
with the UV luminosity when the several brightest galaxies are excluded. 
The LAEs do not display significantly bluer slopes than the LBGs in this 
sample and those in previous studies, suggesting that most LAEs are probably
not younger than LBGs as expected previously. 
A small fraction of our galaxies have extremely steep 
UV slopes with $\beta\simeq-3$. Stellar populations in these galaxies could be 
very young with extremely low metallicity and dust content. 
Our galaxies have moderately strong \lya\ EWs in a wide range of $\sim$10 to 
$\sim$200 \AA. The SFRs estimated from the \lya\ and UV luminosities are also 
moderate, from a few to a few tens $M_{\sun}$ yr$^{-1}$. 

We have also derived the UV LFs of LAEs with $\rm EW>20$ \AA\ at $z\simeq5.7$ 
and 6.5, using our deep $HST$ near-IR data. We have confirmed that the number 
density of LAEs ($\rm EW>20$ \AA) is comparable to that of LBGs at the bright 
end. This suggests that the LAE fraction among bright LBGs is very high.
We also concluded that the number density of LAEs cannot be as high as
that of LBGs at the faint end, even if we maximize the former
by adjusting a model parameter $b$ (i.e., $b=0$). This implies that there
exists a substantial population of faint LBGs with weak \lya\ emission
($\rm EW<20$ \AA) that could be the dominant contribution to the total
UV luminosity density at $z\ge6$.

The LAEs and LBGs in this sample are indistinguishable in many aspects
of the \lya\ and UV continuum properties. In fact, in the 
EW(\lya)--$M_{\rm UV}$ diagram, the distributions of the two populations are
quite similar, showing no sign of enhanced Ly$\alpha$ strengths among
the LAE population. All these seem to indicate that LAEs can be considered
as a subset of LBGs with strong \lya\ emission lines.

\acknowledgments

Support for this work was provided by NASA through Hubble Fellowship grant 
HST-HF-51291.01 awarded by the Space Telescope Science Institute (STScI), 
which is operated by the Association of Universities for Research in
Astronomy, Inc., for NASA, under contract NAS 5-26555.
L.J., E.E., M.M, and S.C. also acknowledge the support from NASA through 
awards issued by STScI ($HST$ PID: 11149,12329,12616) and by JPL/Caltech 
($Spitzer$ PID: 40026,70094). We thank Richard S. Ellis and Daniel P. Stark
for valuable discussions. 

{\it Facilities:}
\facility{$HST$ (NICMOS,WFC3)},
\facility{$Spitzer$ (IRAC)},
\facility{$Subaru$ (Suprime-Cam)}

\appendix
\section{THUMBNAIL IMAGES OF THE GALAXIES}
Thumbnail images of the galaxies in one of the three Subaru Suprime-Cam narrow 
bands (NB816, NB921, and NB973, depending on redshift), three Suprime-Cam 
broad bands $iz'y$, two $HST$ near-IR bands $J_{125}$ (or $J_{110}$) and 
$H_{160}$, and two $Spitzer$ IRAC bands. The galaxies are in the middle of the 
images. The size of the images is $6\farcs6 \times 6\farcs6$
(north is up and east to the left).

\clearpage
\begin{deluxetable}{ccccccccccccc}
\tablecaption{Optical and Near-IR Photometry of the Galaxies}
\tablewidth{0pt}
\tablehead{\colhead{No.} & \colhead{R.A.} & \colhead{Decl.} &
	\colhead{Redshift} & \colhead{NB\tablenotemark{a}} & \colhead{$z'$} & 
	\colhead{$y$} & \colhead{F110W} & \colhead{F125W} & \colhead{F160W} &
	\colhead{Ref.\tablenotemark{b}} \\
	\colhead{} & \colhead{(J2000.0)} & \colhead{(J2000.0)} & \colhead{} & 
	\colhead{(mag)} & \colhead{(mag)} & \colhead{(mag)} & \colhead{(mag)} & 
	\colhead{(mag)} & \colhead{(mag)} }
\startdata
 1& 13:24:38.940& +27:13:40.95& 5.645& 25.88$\pm$0.15&    $\ldots$   &    $\ldots$   &    $\ldots$   &    $\ldots$   &    $\ldots$    & 1 \\
 2& 13:23:54.601& +27:24:12.72& 5.654& 25.01$\pm$0.07& 26.46$\pm$0.14& 26.29$\pm$0.27& 27.11$\pm$0.10& 27.05$\pm$0.12& 27.13$\pm$0.13 & 1 \\
 3& 13:24:16.468& +27:19:07.65& 5.665& 24.35$\pm$0.04& 25.10$\pm$0.04& 25.03$\pm$0.08& 25.24$\pm$0.04&    $\ldots$   & 25.26$\pm$0.07 & 1 \\
 4& 13:24:32.885& +27:30:08.82& 5.671& 24.75$\pm$0.04& 25.29$\pm$0.05& 25.31$\pm$0.14&    $\ldots$   & 26.09$\pm$0.11& 26.08$\pm$0.14 & 2 \\
 5& 13:24:11.887& +27:41:31.81& 5.681& 24.65$\pm$0.05& 26.40$\pm$0.13& 26.62$\pm$0.36&    $\ldots$   & 26.81$\pm$0.11& 26.95$\pm$0.17 & 1 \\
 6& 13:24:11.868& +27:19:48.23& 5.682& 25.41$\pm$0.09&    $\ldots$   &    $\ldots$   &    $\ldots$   & 27.90$\pm$0.18&    $\ldots$    & 1 \\
 7& 13:24:15.987& +27:16:11.05& 5.691& 24.74$\pm$0.05& 26.37$\pm$0.11& 26.74$\pm$0.27&    $\ldots$   & 26.66$\pm$0.12& 27.06$\pm$0.20 & 1 \\
 8& 13:23:47.120& +27:24:13.82& 5.692& 26.10$\pm$0.17&    $\ldots$   &    $\ldots$   &    $\ldots$   &    $\ldots$   &    $\ldots$    & 2 \\
 9& 13:24:28.313& +27:30:12.17& 5.693& 25.41$\pm$0.09&    $\ldots$   &    $\ldots$   &    $\ldots$   &    $\ldots$   &    $\ldots$    & 2 \\
10& 13:24:33.097& +27:29:38.58& 5.696& 25.69$\pm$0.11& 26.92$\pm$0.20& 26.63$\pm$0.35&    $\ldots$   & 27.11$\pm$0.13& 27.40$\pm$0.23 & 1 \\
11& 13:25:20.192& +27:18:42.27& 5.697& 25.91$\pm$0.16&    $\ldots$   &    $\ldots$   &    $\ldots$   &    $\ldots$   &    $\ldots$    & 2 \\
12& 13:24:16.128& +27:44:11.62& 5.698& 24.11$\pm$0.04& 24.71$\pm$0.04& 24.28$\pm$0.04& 24.26$\pm$0.03&    $\ldots$   & 24.23$\pm$0.04 & 1 \\
13& 13:24:18.082& +27:16:38.93& 5.698& 25.83$\pm$0.13&    $\ldots$   &    $\ldots$   &    $\ldots$   &    $\ldots$   &    $\ldots$    & 2 \\
14& 13:25:23.411& +27:17:01.34& 5.705& 24.66$\pm$0.06& 25.33$\pm$0.05& 25.32$\pm$0.13& 25.36$\pm$0.10&    $\ldots$   & 25.47$\pm$0.19 & 1 \\
15& 13:24:23.705& +27:33:24.82& 5.710& 23.57$\pm$0.03& 24.67$\pm$0.03& 24.74$\pm$0.06& 24.65$\pm$0.03&    $\ldots$   & 24.87$\pm$0.06 & 1 \\
16& 13:24:13.004& +27:41:45.80& 5.715& 25.90$\pm$0.15&    $\ldots$   &    $\ldots$   &    $\ldots$   &    $\ldots$   &    $\ldots$    & 1 \\
17& 13:23:44.747& +27:24:26.81& 5.716& 25.64$\pm$0.11&    $\ldots$   &    $\ldots$   &    $\ldots$   & 26.86$\pm$0.13& 27.09$\pm$0.21 & 2 \\
18& 13:24:37.191& +27:35:02.36& 5.717& 25.60$\pm$0.11&    $\ldots$   &    $\ldots$   &    $\ldots$   &    $\ldots$   &    $\ldots$    & 1 \\
19& 13:25:22.120& +27:35:46.87& 5.718& 25.41$\pm$0.09& 26.20$\pm$0.10& 26.64$\pm$0.36&    $\ldots$   & 25.98$\pm$0.12& 26.03$\pm$0.16 & 1 \\
20& 13:24:40.527& +27:13:57.91& 5.724& 24.10$\pm$0.04& 26.21$\pm$0.11& 26.27$\pm$0.27&    $\ldots$   & 26.04$\pm$0.09& 26.05$\pm$0.11 & 1 \\
21& 13:24:30.633& +27:29:34.61& 5.738& 25.64$\pm$0.11& 26.59$\pm$0.13&    $\ldots$   &    $\ldots$   & 27.17$\pm$0.14& 27.19$\pm$0.18 & 1 \\
22& 13:24:41.264& +27:26:49.09& 5.743& 25.63$\pm$0.10& 26.69$\pm$0.16&    $\ldots$   &    $\ldots$   & 26.81$\pm$0.17& 26.81$\pm$0.21 & 2 \\
23& 13:24:18.450& +27:16:32.56& 5.922&     $\ldots$  & 25.43$\pm$0.05& 25.66$\pm$0.14&    $\ldots$   & 25.58$\pm$0.05& 25.64$\pm$0.07 & 3 \\
24& 13:25:19.463& +27:18:28.51& 6.002&     $\ldots$  & 25.40$\pm$0.05& 25.55$\pm$0.14& 25.59$\pm$0.08&    $\ldots$   & 25.70$\pm$0.14 & 4 \\
25& 13:24:26.559& +27:15:59.72& 6.032&     $\ldots$  & 25.24$\pm$0.05& 25.28$\pm$0.10& 25.86$\pm$0.12&    $\ldots$   & 25.60$\pm$0.11 & 5 \\
26& 13:24:31.551& +27:15:08.72& 6.032&     $\ldots$  & 25.85$\pm$0.09&    $\ldots$   &    $\ldots$   &    $\ldots$   &    $\ldots$    & 3 \\
27& 13:24:10.766& +27:19:03.95& 6.040&     $\ldots$  & 26.56$\pm$0.15&    $\ldots$   &    $\ldots$   & 26.42$\pm$0.11& 26.81$\pm$0.19 & 6 \\
28& 13:24:42.452& +27:24:23.35& 6.042&     $\ldots$  & 25.75$\pm$0.07&    $\ldots$   & 25.84$\pm$0.10&    $\ldots$   & 26.74$\pm$0.35 & 5 \\
29& 13:24:05.895& +27:18:37.72& 6.049&     $\ldots$  & 27.00$\pm$0.20&    $\ldots$   &    $\ldots$   & 27.27$\pm$0.13& 27.35$\pm$0.18 & 6 \\
30& 13:24:00.301& +27:32:37.95& 6.062&     $\ldots$  & 25.78$\pm$0.07& 26.11$\pm$0.23& 26.11$\pm$0.13&    $\ldots$   & 26.69$\pm$0.26 & $\ast$ \\
31& 13:23:45.632& +27:17:00.53& 6.112&     $\ldots$  & 25.09$\pm$0.04&    $\ldots$   & 26.24$\pm$0.14&    $\ldots$   & 26.40$\pm$0.19 & 4 \\
32& 13:24:55.583& +27:39:20.89& 6.127&     $\ldots$  & 26.79$\pm$0.19&    $\ldots$   &    $\ldots$   &    $\ldots$   &    $\ldots$    & 6 \\
33& 13:24:20.628& +27:16:40.47& 6.269&     $\ldots$  & 26.63$\pm$0.16&    $\ldots$   &    $\ldots$   & 26.79$\pm$0.13& 26.88$\pm$0.17 & 6 \\
34& 13:23:45.757& +27:32:51.30& 6.315&     $\ldots$  & 25.55$\pm$0.06& 25.04$\pm$0.08&    $\ldots$   & 24.90$\pm$0.04& 24.90$\pm$0.04 & 6 \\
35& 13:24:40.643& +27:36:06.94& 6.332&     $\ldots$  & 25.55$\pm$0.06&    $\ldots$   &    $\ldots$   & 25.45$\pm$0.06& 25.52$\pm$0.07 & 7 \\
36& 13:23:45.937& +27:25:18.06& 6.482&     $\ldots$  & 25.71$\pm$0.07& 25.34$\pm$0.12&    $\ldots$   & 25.75$\pm$0.07& 25.69$\pm$0.08 & $\ast$ \\
37& 13:24:18.416& +27:33:44.97& 6.508& 25.09$\pm$0.06& 26.69$\pm$0.17& 26.56$\pm$0.34& 26.77$\pm$0.16&    $\ldots$   & 26.82$\pm$0.28 & 8 \\
38& 13:24:34.284& +27:40:56.32& 6.519& 25.71$\pm$0.12&    $\ldots$   &    $\ldots$   &    $\ldots$   &    $\ldots$   &    $\ldots$    & 2 \\
39& 13:23:43.190& +27:24:52.04& 6.534& 24.44$\pm$0.04& 26.77$\pm$0.20&    $\ldots$   &    $\ldots$   & 26.59$\pm$0.13& 26.52$\pm$0.15 & 2 \\
40& 13:24:55.772& +27:40:15.31& 6.534& 25.82$\pm$0.11& 27.39$\pm$0.33& 26.13$\pm$0.24&    $\ldots$   & 26.69$\pm$0.12& 27.25$\pm$0.25 & 2 \\
41& 13:24:58.508& +27:39:12.92& 6.537& 25.49$\pm$0.08&    $\ldots$   &    $\ldots$   &    $\ldots$   &    $\ldots$   &    $\ldots$    & 9 \\
42& 13:23:49.186& +27:32:11.39& 6.539& 25.60$\pm$0.09&    $\ldots$   &    $\ldots$   &    $\ldots$   &    $\ldots$   &    $\ldots$    & 2 \\
43& 13:23:53.054& +27:16:30.75& 6.542& 25.18$\pm$0.06& 26.38$\pm$0.13& 26.04$\pm$0.21&    $\ldots$   & 26.23$\pm$0.08& 26.50$\pm$0.12 & 8 \\
44& 13:24:15.678& +27:30:57.79& 6.543& 23.92$\pm$0.03& 25.51$\pm$0.05& 25.10$\pm$0.05&    $\ldots$   & 25.12$\pm$0.05& 24.99$\pm$0.06 & 8,11 \\
45& 13:24:40.239& +27:25:53.11& 6.544& 25.70$\pm$0.10&    $\ldots$   &    $\ldots$   &    $\ldots$   & 26.98$\pm$0.19& 27.28$\pm$0.30 & 2 \\
46& 13:23:52.680& +27:16:21.76& 6.545& 25.59$\pm$0.09&    $\ldots$   &    $\ldots$   &    $\ldots$   & 26.69$\pm$0.13& 26.66$\pm$0.16 & 8 \\
47& 13:24:10.817& +27:19:28.08& 6.547& 24.66$\pm$0.07& 26.19$\pm$0.10& 25.02$\pm$0.08&    $\ldots$   & 24.65$\pm$0.03& 24.67$\pm$0.04 & 9 \\
48& 13:23:48.922& +27:15:30.33& 6.548& 25.45$\pm$0.08&    $\ldots$   &    $\ldots$   &    $\ldots$   & 27.20$\pm$0.19& 27.22$\pm$0.25 & 2 \\
49& 13:24:17.909& +27:17:45.94& 6.548& 25.19$\pm$0.07& 27.23$\pm$0.30& 26.43$\pm$0.32&    $\ldots$   & 26.32$\pm$0.10& 26.35$\pm$0.13 & 2 \\
50& 13:23:44.896& +27:31:44.90& 6.550& 24.82$\pm$0.05& 26.36$\pm$0.13& 26.27$\pm$0.23&    $\ldots$   & 26.70$\pm$0.12& 26.61$\pm$0.14 & 2 \\
51& 13:23:47.710& +27:23:59.63& 6.553& 25.47$\pm$0.08&    $\ldots$   &    $\ldots$   &    $\ldots$   & 27.05$\pm$0.19&    $\ldots$    & 2 \\
52& 13:24:35.005& +27:39:57.43& 6.554& 25.28$\pm$0.07& 27.32$\pm$0.31&    $\ldots$   &    $\ldots$   & 27.57$\pm$0.15& 27.67$\pm$0.21 & 2 \\
53& 13:24:28.652& +27:30:49.24& 6.555& 25.58$\pm$0.10&    $\ldots$   &    $\ldots$   &    $\ldots$   &    $\ldots$   &    $\ldots$    & 2 \\
54& 13:24:08.313& +27:15:43.49& 6.556& 24.34$\pm$0.04& 25.87$\pm$0.08& 25.55$\pm$0.13&    $\ldots$   & 25.96$\pm$0.14& 25.82$\pm$0.15 & 8 \\
55& 13:24:19.326& +27:41:24.82& 6.564& 25.31$\pm$0.08&    $\ldots$   &    $\ldots$   &    $\ldots$   & 27.94$\pm$0.19&    $\ldots$    & 2 \\
56& 13:25:23.326& +27:16:12.44& 6.567& 25.17$\pm$0.08&    $\ldots$   &    $\ldots$   & 26.85$\pm$0.17&    $\ldots$   &    $\ldots$    & 2 \\
57& 13:25:18.763& +27:30:43.45& 6.580& 25.14$\pm$0.06&    $\ldots$   &    $\ldots$   &    $\ldots$   &    $\ldots$   &    $\ldots$    & 8 \\
58& 13:24:43.427& +27:26:32.62& 6.583& 25.30$\pm$0.07& 26.62$\pm$0.15& 25.74$\pm$0.17&    $\ldots$   & 25.77$\pm$0.09& 25.69$\pm$0.10 & 2 \\
59& 13:23:57.128& +27:24:47.63& 6.585& 25.46$\pm$0.08& 26.93$\pm$0.21&    $\ldots$   & 27.43$\pm$0.19&    $\ldots$   & 27.25$\pm$0.21 & 9 \\
60& 13:25:20.446& +27:34:59.29& 6.592& 25.36$\pm$0.07&    $\ldots$   &    $\ldots$   &    $\ldots$   &    $\ldots$   &    $\ldots$    & 9 \\
61& 13:25:22.291& +27:35:19.95& 6.599& 24.71$\pm$0.04& 26.49$\pm$0.13& 25.57$\pm$0.10&    $\ldots$   & 25.93$\pm$0.14& 26.26$\pm$0.24 & 8 \\
62& 13:23:59.766& +27:24:55.75& 6.964& 24.57$\pm$0.08& 26.99$\pm$0.29& 24.91$\pm$0.09& 25.50$\pm$0.04& 25.42$\pm$0.05& 25.44$\pm$0.05 & 10 \\
63& 02:18:00.899& -05:11:37.69& 6.023&     $\ldots$  & 24.91$\pm$0.06& 25.01$\pm$0.15&    $\ldots$   & 25.16$\pm$0.06& 25.38$\pm$0.09 & 12 \\
64& 02:17:35.337& -05:10:32.50& 6.116&     $\ldots$  & 25.35$\pm$0.09& 25.22$\pm$0.18&    $\ldots$   & 25.22$\pm$0.07& 25.56$\pm$0.09 & 12 \\
65& 02:18:19.420& -05:09:00.57& 6.563& 25.30$\pm$0.12&    $\ldots$   &    $\ldots$   &    $\ldots$   &    $\ldots$   &    $\ldots$    & 13 \\
66& 02:18:20.701& -05:11:09.89& 6.575& 24.97$\pm$0.09&    $\ldots$   &    $\ldots$   &    $\ldots$   & 26.72$\pm$0.20& 26.68$\pm$0.23 & 13 \\
67& 02:17:57.585& -05:08:44.72& 6.595& 23.99$\pm$0.04& 25.82$\pm$0.15& 25.03$\pm$0.15&    $\ldots$   & 24.61$\pm$0.08& 24.80$\pm$0.08 & 13,14 \\
\enddata
\tablenotetext{a}{NB = NB816 for $z\simeq5.7$ LAEs, NB = NB921 for 
$z\simeq6.5$ LAEs, NB = NB973 for the $z\simeq7$ LAE.}
\tablenotetext{b}{References: (1) \citet{shi06}; (2) \citet{kas11}; 
(3) \citet{ota08}; (4) \citet{nag07}; (5) \citet{nag05}; (6) \citet{jia11}; 
(7) \citet{nag04}; (8) \citet{tan05}; (9) \citet{kas06}; (10) \citet{iye06}; 
(11) \citet{kod03}; (12) \citet{cur12}; (13) \citet{ouc10}; (14) \citet{ouc09};
($\ast$) unpublished.}
\tablecomments{The first 62 galaxies are from the SDF, and the last 5 
galaxies are from the SXDS. }
\end{deluxetable}

\clearpage
\begin{deluxetable}{ccccccccccccc}
\tablecaption{Spectral Properties of the Galaxies}
\tablewidth{0pt}
\tablehead{\colhead{No.} & \colhead{$M_{1500}$} & \colhead{UV Slope} &
   \colhead{L(\lya)} & \colhead{EW(\lya)} & \colhead{SFR(UV)} & 
	\colhead{SFR(\lya)} \\
   \colhead{} & \colhead{(mag)} & \colhead{($\beta$)} & 
	\colhead{($\rm 10^{42}\ erg\ s^{-1}$)} & \colhead{(\AA)} & 
	\colhead{($\rm M_{\sun}\ yr^{-1}$)} & \colhead{($\rm M_{\sun}\ yr^{-1}$)} }
\startdata
 2 & $-19.53\pm0.15$ & $-2.09\pm0.52$ & $ 13.5\pm 1.2$ & $231.3\pm38.6$ & $  3.8\pm 0.5$ & $ 12.3\pm 1.1$ \\
 3 & $-21.46\pm0.03$ & $-2.36\pm0.13$ & $ 10.5\pm 0.5$ & $ 28.6\pm 1.6$ & $ 20.9\pm 0.5$ & $  9.5\pm 0.5$ \\
 4 & $-20.50\pm0.27$ & $-1.96\pm0.79$ & $  8.7\pm 0.5$ & $ 62.7\pm16.0$ & $  9.9\pm 2.5$ & $  8.0\pm 0.4$ \\
 5 & $-20.07\pm0.09$ & $-3.00\pm0.35$ & $  8.7\pm 0.6$ & $ 75.2\pm 8.0$ & $  4.9\pm 0.4$ & $  7.9\pm 0.5$ \\
 7 & $-20.12\pm0.08$ & $-3.01\pm0.34$ & $  7.3\pm 0.5$ & $ 60.2\pm 5.9$ & $  5.1\pm 0.4$ & $  6.7\pm 0.4$ \\
10 & $-19.69\pm0.13$ & $-2.89\pm0.49$ & $  2.7\pm 0.4$ & $ 33.8\pm 6.3$ & $  3.5\pm 0.4$ & $  2.5\pm 0.4$ \\
14 & $-21.27\pm0.04$ & $-2.19\pm0.28$ & $  6.4\pm 0.5$ & $ 21.6\pm 1.9$ & $ 18.4\pm 0.7$ & $  5.8\pm 0.5$ \\
15 & $-21.94\pm0.02$ & $-2.21\pm0.11$ & $ 20.8\pm 0.8$ & $ 37.8\pm 1.7$ & $ 33.9\pm 0.6$ & $ 19.0\pm 0.7$ \\
17 & $-19.99\pm0.34$ & $-3.02\pm1.10$ & $  3.0\pm 0.4$ & $ 28.0\pm 9.7$ & $  4.5\pm 1.4$ & $  2.8\pm 0.4$ \\
19 & $-20.46\pm0.08$ & $-1.57\pm0.30$ & $  4.0\pm 0.5$ & $ 32.0\pm 4.4$ & $ 10.9\pm 0.8$ & $  3.6\pm 0.4$ \\
20 & $-20.46\pm0.08$ & $-1.68\pm0.26$ & $ 16.5\pm 0.9$ & $131.1\pm11.4$ & $ 10.5\pm 0.7$ & $ 15.1\pm 0.8$ \\
21 & $-19.85\pm0.10$ & $-3.18\pm0.37$ & $  4.4\pm 0.6$ & $ 44.9\pm 7.6$ & $  3.8\pm 0.3$ & $  4.0\pm 0.6$ \\
22 & $-19.90\pm0.12$ & $-2.24\pm0.44$ & $  5.1\pm 0.7$ & $ 60.9\pm10.4$ & $  5.1\pm 0.6$ & $  4.7\pm 0.6$ \\
23 & $-21.17\pm0.03$ & $-2.35\pm0.14$ & $  5.0\pm 0.8$ & $ 18.0\pm 2.8$ & $ 16.0\pm 0.5$ & $  4.6\pm 0.7$ \\
24 & $-21.19\pm0.04$ & $-2.58\pm0.22$ & $ 13.6\pm 0.8$ & $ 45.6\pm 3.1$ & $ 15.2\pm 0.5$ & $ 12.4\pm 0.7$ \\
25 & $-21.30\pm0.04$ & $-2.82\pm0.20$ & $ 14.6\pm 1.2$ & $ 42.0\pm 3.8$ & $ 15.9\pm 0.6$ & $ 13.3\pm 1.1$ \\
26 & $  \ldots     $ & $   \ldots   $ & $ 17.4\pm 1.2$ & $  \ldots    $ & $   \ldots   $ & $ 15.8\pm 1.1$ \\
27 & $-20.20\pm0.09$ & $-2.20\pm0.40$ & $  6.9\pm 0.8$ & $ 59.5\pm 8.6$ & $  6.9\pm 0.6$ & $  6.3\pm 0.7$ \\
28 & $-20.85\pm0.06$ & $-2.81\pm0.39$ & $ 18.3\pm 1.2$ & $ 80.7\pm 7.1$ & $ 10.4\pm 0.6$ & $ 16.6\pm 1.1$ \\
29 & $-19.58\pm0.12$ & $-2.62\pm0.47$ & $  2.9\pm 0.8$ & $ 41.3\pm12.6$ & $  3.4\pm 0.4$ & $  2.6\pm 0.7$ \\
30 & $-20.68\pm0.07$ & $-3.45\pm0.38$ & $  4.6\pm 0.8$ & $ 20.5\pm 3.9$ & $  7.8\pm 0.5$ & $  4.2\pm 0.8$ \\
31 & $-20.53\pm0.19$ & $-2.51\pm0.76$ & $ 18.8\pm 0.4$ & $117.4\pm20.2$ & $  8.4\pm 1.4$ & $ 17.1\pm 0.4$ \\
32 & $  \ldots     $ & $   \ldots   $ & $  5.9\pm 0.8$ & $  \ldots    $ & $   \ldots   $ & $  5.3\pm 0.8$ \\
33 & $-20.03\pm0.24$ & $-2.40\pm0.95$ & $  4.4\pm 0.9$ & $ 44.7\pm13.4$ & $  5.5\pm 1.2$ & $  4.0\pm 0.8$ \\
34 & $-21.82\pm0.04$ & $-1.79\pm0.17$ & $  4.1\pm 0.9$ & $  8.6\pm 1.9$ & $ 35.0\pm 1.4$ & $  3.7\pm 0.8$ \\
35 & $-21.37\pm0.11$ & $-2.31\pm0.41$ & $ 18.1\pm 0.5$ & $ 54.2\pm 5.5$ & $ 19.5\pm 1.9$ & $ 16.5\pm 0.4$ \\
36 & $-21.25\pm0.07$ & $-2.57\pm0.30$ & $  2.8\pm 1.0$ & $  8.9\pm 3.1$ & $ 16.1\pm 1.0$ & $  2.5\pm 0.9$ \\
37 & $-20.19\pm0.26$ & $-2.55\pm0.94$ & $ 12.6\pm 1.0$ & $106.5\pm26.6$ & $  6.1\pm 1.4$ & $ 11.5\pm 0.9$ \\
39 & $-20.21\pm0.20$ & $-1.69\pm0.88$ & $ 16.1\pm 0.8$ & $161.1\pm31.3$ & $  8.3\pm 1.6$ & $ 14.7\pm 0.8$ \\
40 & $-20.39\pm0.12$ & $-4.37\pm0.73$ & $  1.4\pm 0.2$ & $  6.8\pm 1.2$ & $  5.2\pm 0.6$ & $  1.3\pm 0.2$ \\
43 & $-20.70\pm0.09$ & $-3.05\pm0.47$ & $  5.2\pm 0.4$ & $ 24.6\pm 2.9$ & $  8.6\pm 0.7$ & $  4.7\pm 0.4$ \\
44 & $-21.74\pm0.03$ & $-1.79\pm0.16$ & $ 19.9\pm 0.8$ & $ 47.4\pm 2.3$ & $ 32.6\pm 1.0$ & $ 18.1\pm 0.7$ \\
45 & $-19.99\pm0.31$ & $-3.33\pm1.57$ & $  3.4\pm 0.4$ & $ 28.9\pm 9.2$ & $  4.2\pm 1.2$ & $  3.1\pm 0.4$ \\
46 & $-20.13\pm0.21$ & $-1.87\pm0.91$ & $  4.1\pm 0.5$ & $ 42.7\pm 9.5$ & $  7.2\pm 1.4$ & $  3.8\pm 0.4$ \\
47 & $-22.08\pm0.03$ & $-1.57\pm0.17$ & $  5.4\pm 0.5$ & $  9.8\pm 0.9$ & $ 48.4\pm 1.5$ & $  4.9\pm 0.4$ \\
48 & $-19.64\pm0.30$ & $-2.09\pm1.39$ & $  5.2\pm 0.5$ & $ 80.8\pm24.2$ & $  4.3\pm 1.2$ & $  4.8\pm 0.5$ \\
49 & $-20.49\pm0.13$ & $-1.98\pm0.59$ & $  5.9\pm 0.5$ & $ 42.5\pm 6.3$ & $  9.7\pm 1.1$ & $  5.4\pm 0.5$ \\
50 & $-20.09\pm0.19$ & $-1.60\pm0.82$ & $  9.6\pm 0.6$ & $108.7\pm20.2$ & $  7.7\pm 1.3$ & $  8.7\pm 0.6$ \\
52 & $-19.31\pm0.24$ & $-2.44\pm1.14$ & $  6.3\pm 0.6$ & $121.6\pm29.1$ & $  2.8\pm 0.6$ & $  5.7\pm 0.5$ \\
54 & $-21.12\pm0.09$ & $-2.64\pm0.42$ & $ 13.0\pm 0.7$ & $ 45.6\pm 4.3$ & $ 14.1\pm 1.1$ & $ 11.8\pm 0.6$ \\
58 & $-21.08\pm0.09$ & $-1.84\pm0.40$ & $  5.3\pm 0.5$ & $ 23.1\pm 2.8$ & $ 17.5\pm 1.4$ & $  4.9\pm 0.4$ \\
61 & $-21.06\pm0.08$ & $-3.47\pm0.48$ & $ 12.6\pm 0.7$ & $ 39.3\pm 3.5$ & $ 11.0\pm 0.8$ & $ 11.4\pm 0.6$ \\
62 & $-21.51\pm0.06$ & $-2.09\pm0.31$ & $ 15.1\pm 1.6$ & $ 41.5\pm 5.0$ & $ 23.9\pm 1.4$ & $ 13.7\pm 1.4$ \\
63 & $-21.67\pm0.04$ & $-2.78\pm0.17$ & $ 18.2\pm 3.0$ & $ 37.7\pm 6.3$ & $ 22.4\pm 0.8$ & $ 16.5\pm 2.7$ \\
64 & $-21.62\pm0.10$ & $-3.01\pm0.38$ & $ 13.0\pm 1.1$ & $ 26.6\pm 3.3$ & $ 20.2\pm 1.9$ & $ 11.8\pm 1.0$ \\
66 & $-20.10\pm0.31$ & $-1.82\pm1.35$ & $  8.3\pm 1.0$ & $ 89.2\pm27.5$ & $  7.1\pm 2.0$ & $  7.6\pm 0.9$ \\
67 & $-22.08\pm0.08$ & $-1.87\pm0.33$ & $ 22.9\pm 1.2$ & $ 39.3\pm 3.5$ & $ 43.3\pm 3.1$ & $ 20.8\pm 1.1$ \\
\enddata
\end{deluxetable}

\clearpage
\begin{figure}	
\epsscale{0.7}
\plotone{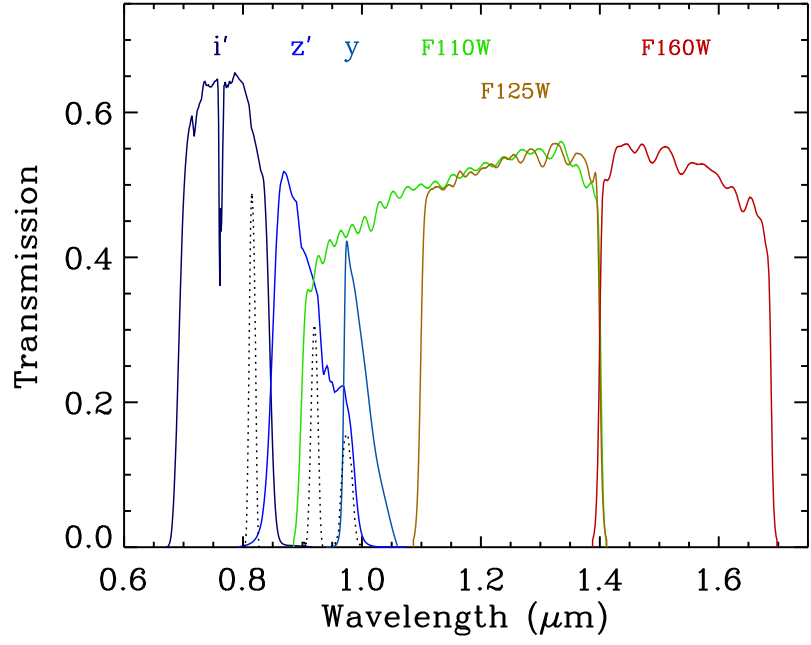}
\caption{Transmission curves of six Subaru Suprime-Cam filters ($i'$, $z'$, 
$y$, NB816, NB921, and NB973) and three $HST$ WFC3 filters (F110W, F125W, 
and F160W). System responses such as detector quantum efficiencies are 
included. Note three narrow-band filters NB816, NB921, and NB973 (dotted
black curves) at $0.8\sim1.0$ $\mu$m.}
\end{figure}

\clearpage
\begin{figure}	
\epsscale{0.7}
\plotone{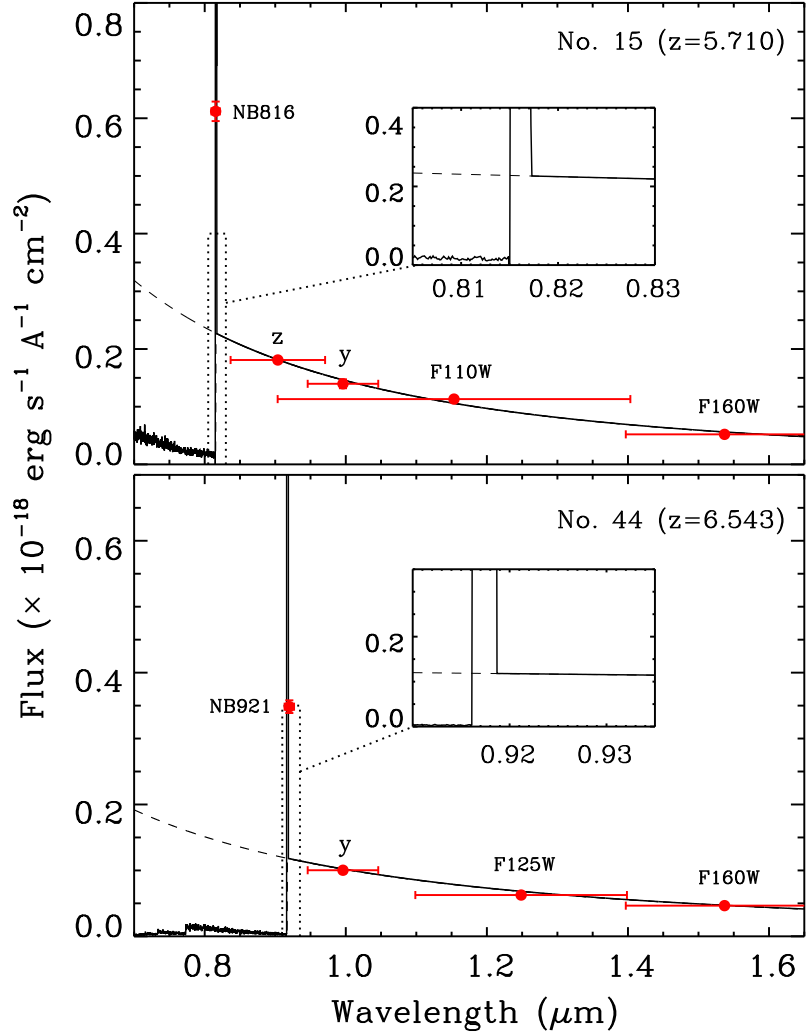}
\caption{Examples of measuring UV continuum and \lya\ line emission. The 
upper panel illustrates our model fit to a LAE at $z=5.710$. The red circles
are the photometric points that we use. The broad-band data in the
$z'yJ_{110}H_{160}$ bands are used to fit the power-law continuum (see Eq. 1).
The horizontal bars indicate the wavelength coverage of the broad bands.
The photometric error bars are also shown, but the errors are very small.
When IGM absorption is applied, the continuum flux at the blue side of the 
\lya\ line is absorbed (the dashed line). The region around the \lya\ emission 
line (the dotted line) is zoomed in to show this absorption. Finally the 
factor $A$ in Eq. 1 is calculated by scaling the \lya\ emission line to match 
the NB816-band photometry. 
The solid profile is the final model spectrum (continuum + \lya) 
for this galaxy. The lower panel shows our model fit to a LAE at $z=6.543$.
In this case we use the data in the $yJ_{125}H_{160}$ bands to compute its
continuum, and use the NB921-band photometry to measure its \lya\ emission.}
\end{figure}

\clearpage
\begin{figure} 
\epsscale{1.0}
\plottwo{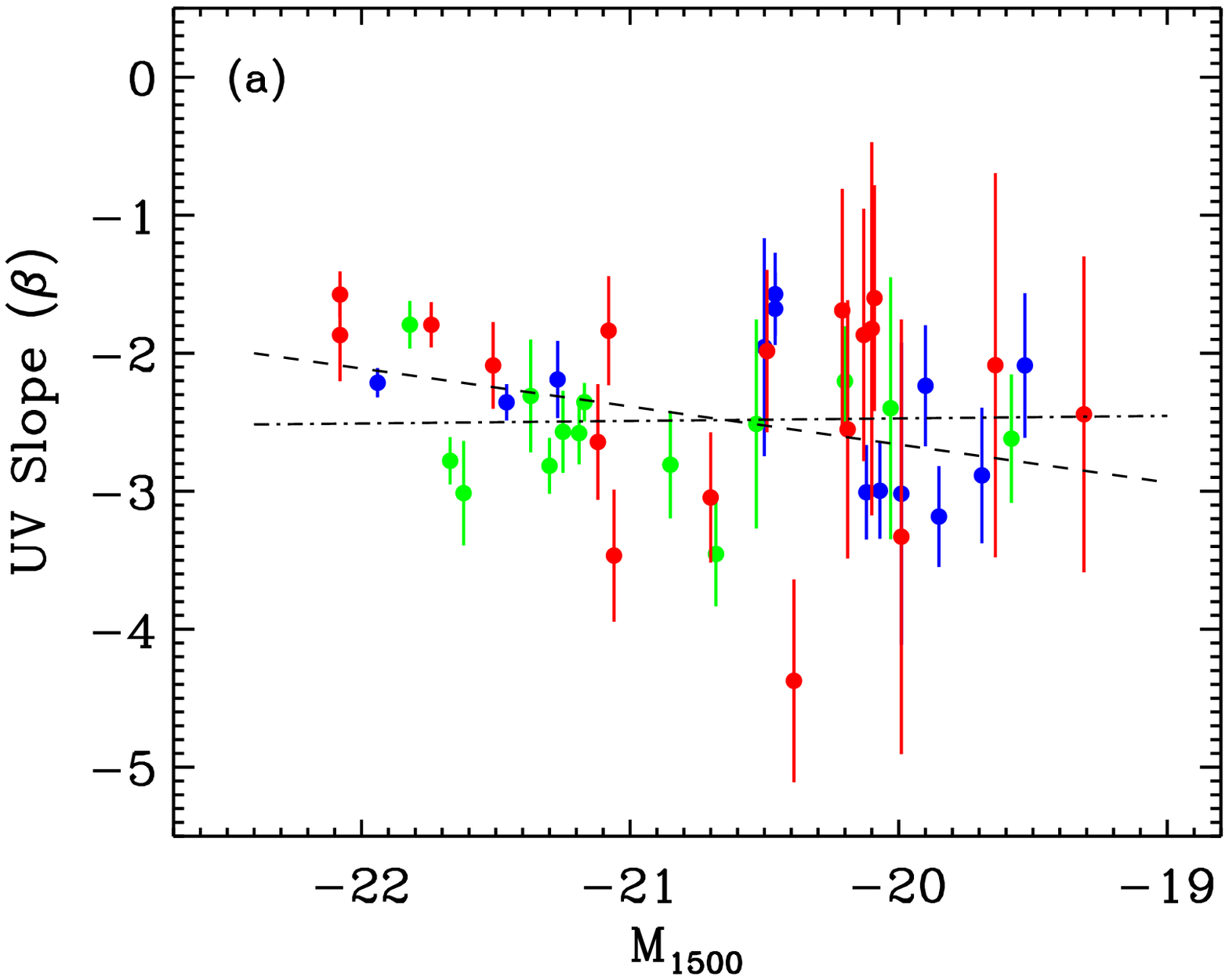}{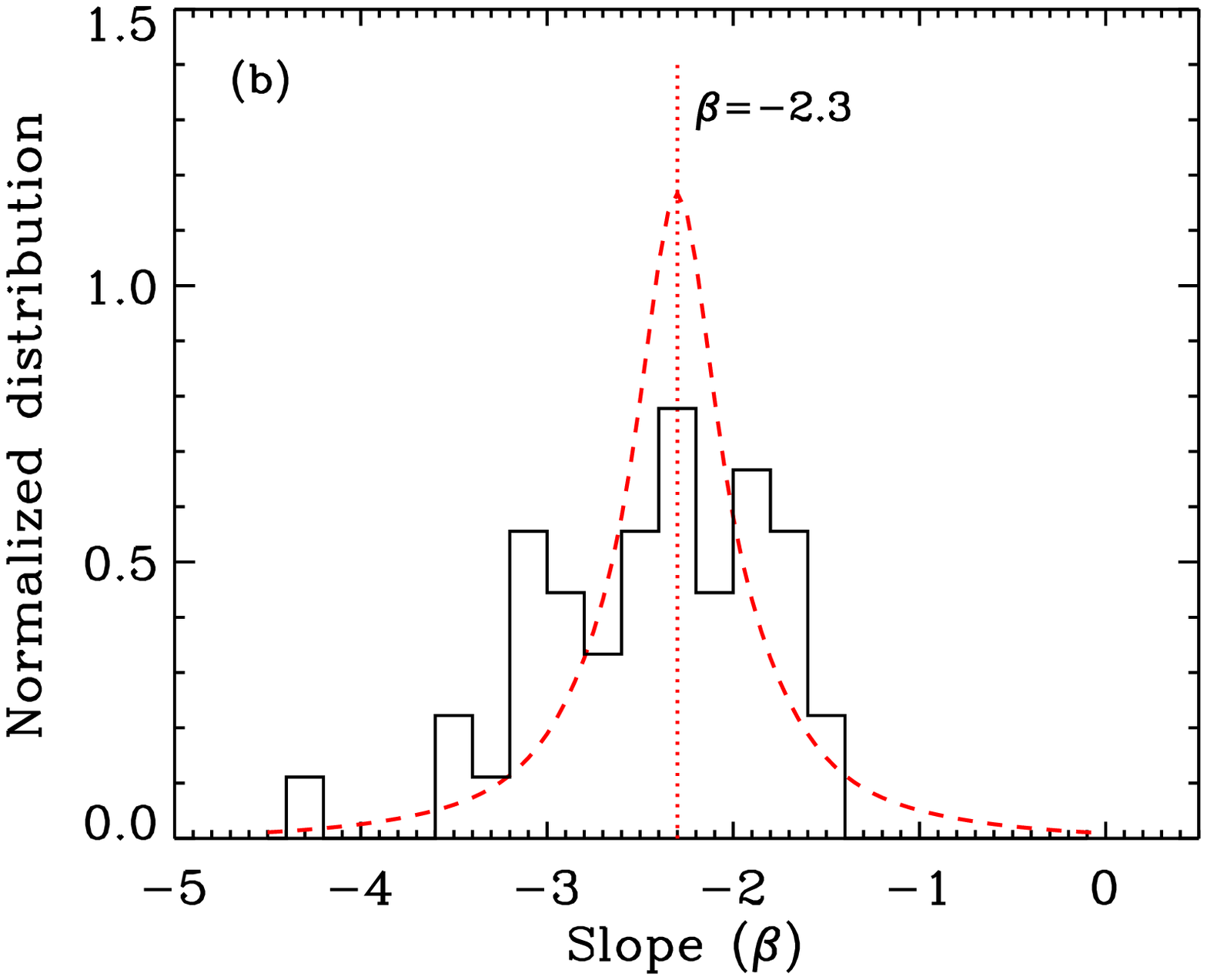}
\caption{(a) UV continuum slope $\beta$ as a function of the UV continuum 
luminosity $M_{1500}$. The blue and red circles represent the LAEs at 
$z\simeq5.7$ and 6.5 (including $z\simeq7$), respectively, and the green 
circles represent the LBGs at $z\simeq6$. The slopes in this sample have a 
weighted mean $\beta\simeq-2.3$. The 
dashed line is the best linear fit to all the data points. The dash-dotted 
line is the best fit to the galaxies fainter than $M_{1500}=-21.7$ mag, which 
suggests that the slope $\beta$ shows little trend with $M_{1500}$ when the 
several brightest galaxies at $M_{1500}\simeq-22$ mag are excluded.
(b) Normalized distribution of the observed $\beta$ (histogram) compared
with the distribution of $\beta$ expected (the red profile) if all the galaxies
have an intrinsic $\beta=-2.3$ (the vertical dotted line) with the observed
uncertainties. The expected $\beta$ distribution is the average of the 
Gaussian distributions for individual galaxies. 
The observed distribution is broader and flatter,
meaning significant intrinsic scatter of $\beta$. It clearly shows a
statistically significant excess of galaxies with $\beta\simeq-3$, indicating
that the existence of $\beta\sim-3$ galaxies are statistically robust.}
\end{figure}

\begin{figure} 
\epsscale{0.7}
\plotone{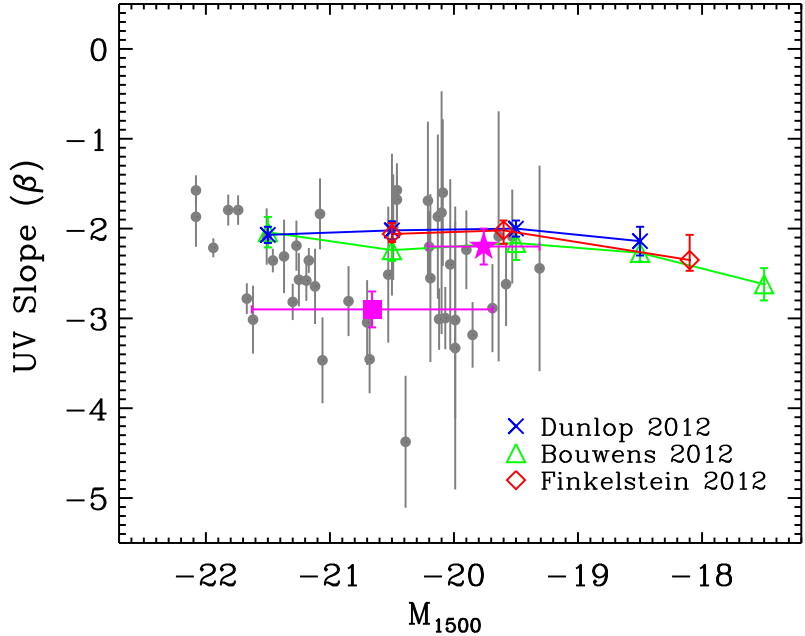}
\caption{UV continuum slope $\beta$ as a function of the UV continuum 
luminosity $M_{1500}$, compared to previous studies. The gray points with 
error bars represent the galaxies shown in Figure 3(a). The magenta star 
represents the stacked image of the 9 objects with $\sigma_\beta>0.8$
that have the $J_{125}$ and $H_{160}$-band images.
The magenta square represents the stacked image of 10 galaxies with 
$\beta<-2.8$ that have the $J_{125}$ and $H_{160}$-band images. The 
horizontal bars on the magenta symbols indicate their magnitude ranges. The 
blue crosses, green triangles, and red diamonds display the average slopes 
from previous studies: the blue crosses are the average $\beta$ at $z=5-7$ 
from \citet{dun12a}; the green triangles are the average $\beta$ at $z\sim6$ 
from \citet{bou12b}; the red diamonds are the average $\beta$ at $z\sim6$ 
from \citet{fin12}. These points are connected by solid lines to guide eyes.
At $M_{1500}\le-19.5$ mag, these studies do not show evidence of a significant 
correlation between $M_{1500}$ and $\beta$. Such relation, if exists, 
could be driven by very faint galaxies.}
\end{figure}

\clearpage
\begin{figure}	
\epsscale{0.6}
\plotone{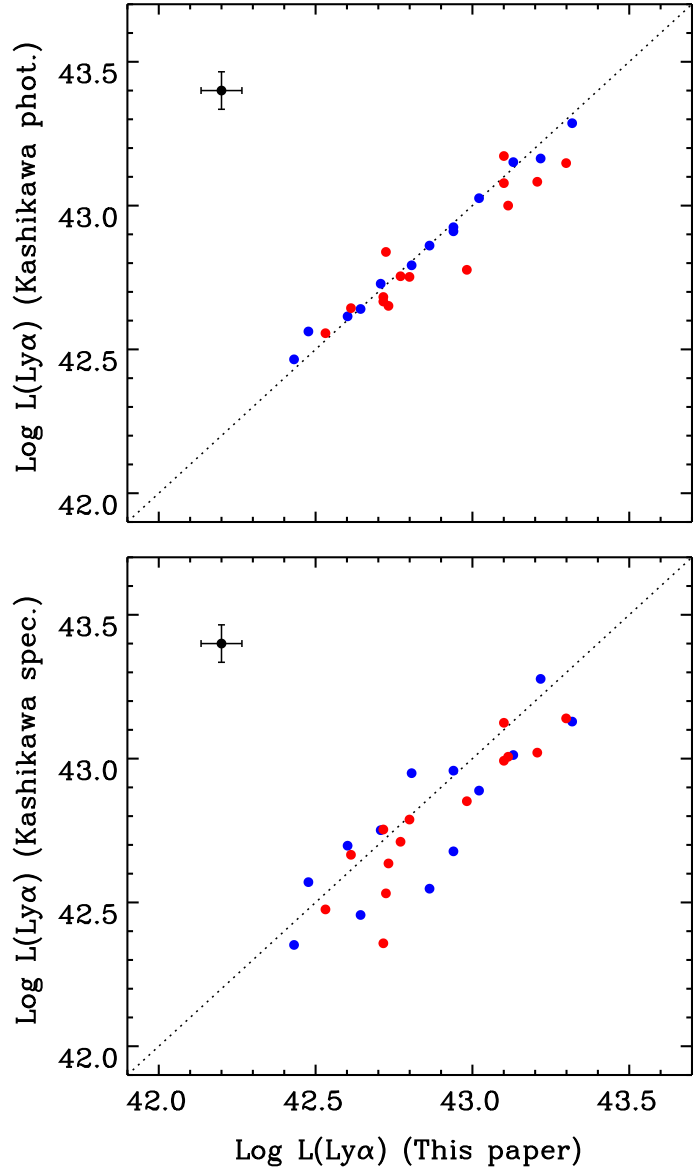}
\caption{Comparison between the observed \lya\ luminosities of the LAEs in 
our paper
and the \lya\ luminosities given in \citet{kas11}. The upper panel compares 
our luminosities to those derived from the photometric data in the galaxy
discovery papers. The photometric data used in these papers are usually the 
narrow-band and $z'$-band photometry. The blue and red circles represent the 
LAEs at $z\simeq5.7$ and 6.5, respectively. The dotted line is the locus where 
the two measurements are equal. The error bars are typical measurement errors.
The two measurements agree well with each other.
The lower panel shows the comparison between our results and the results from 
the spectroscopic data in \citet{kas11}. The two are also consistent, with a
slightly larger scatter than that in the upper panel.}
\end{figure}

\clearpage
\begin{figure}	
\epsscale{0.7}
\plotone{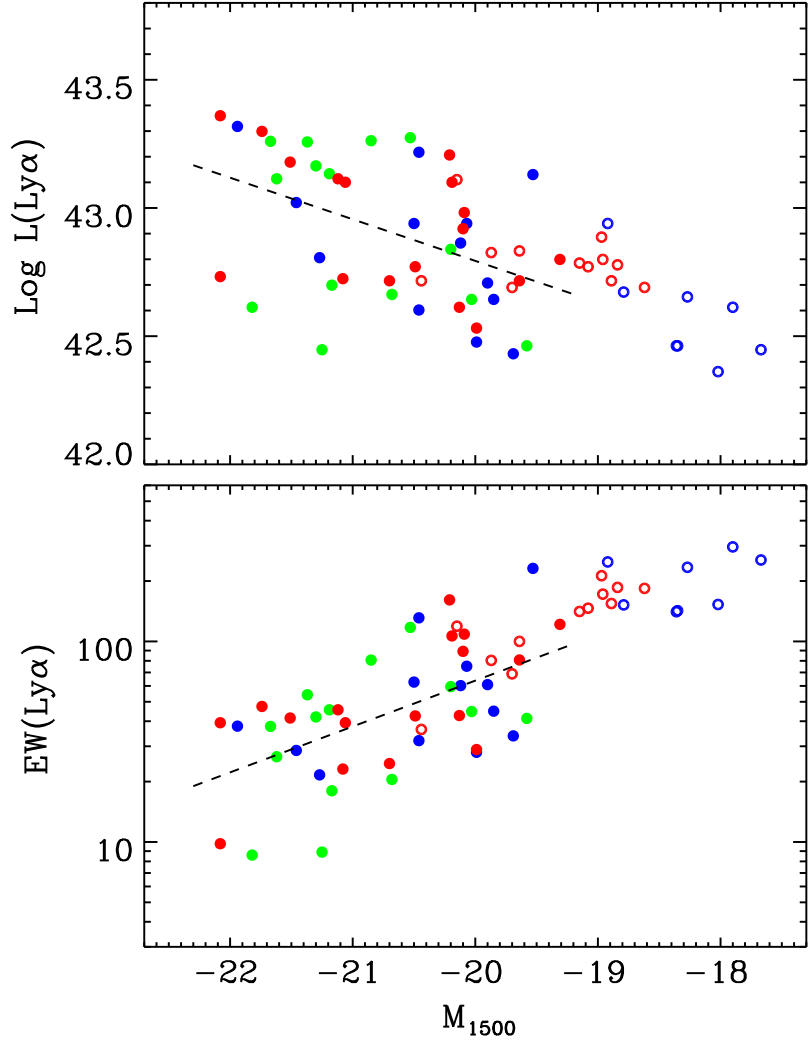}
\caption{\lya\ luminosity and rest-frame EW (\AA) 
as a function of UV luminosity 
$M_{1500}$. The color-coded filled circles have the same meaning as in Figure 
3. The open circles represent the LAEs with weak detections ($<5\sigma$) in 
the $J$ band, and their measurements of physical quantities are crude (see 
Section 4.2). The dashed lines are the best linear fit to the filled 
circles. The \lya\ luminosity shows weak dependence on $M_{1500}$ in the upper
panel. The lower panel shows that lower-luminosity galaxies tend to have 
higher \lya\ EWs. The real relation should be weaker than it appears
due to the nature of the flux-limited sample.
There is no significant difference between LAEs and LBGs in 
the range of $M_{1500}<-19.5$ mag.}
\end{figure}

\clearpage
\begin{figure}	
\epsscale{0.7}
\plotone{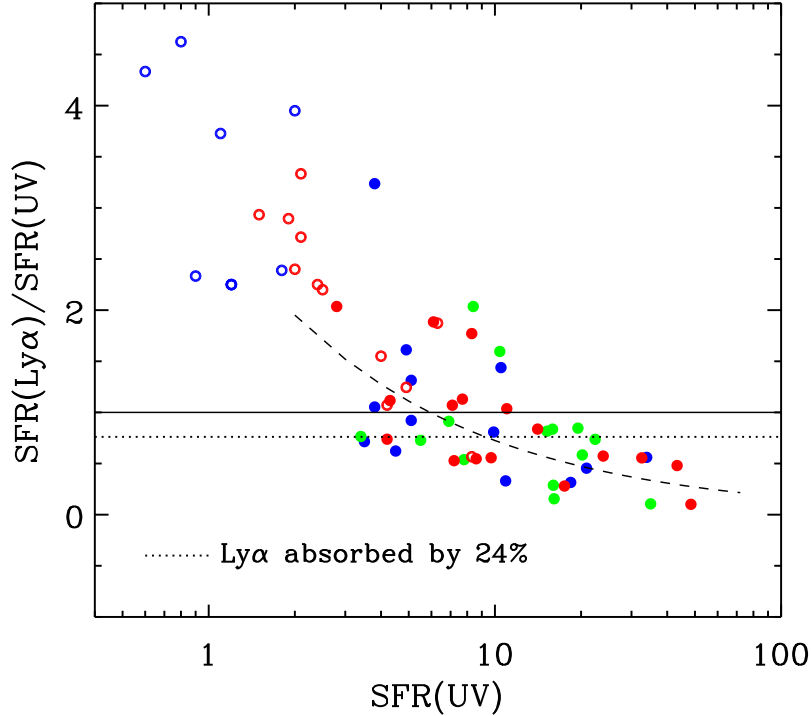}
\caption{SFRs estimated from UV continuum luminosities and \lya\ luminosities 
in units of $M_{\sun}$ yr$^{-1}$. The color-coded symbols have the same 
meaning as in Figure 6. The solid line represents the position where the two 
measurements are equal. The dotted line is the expected ratio of the \lya\ 
SFRs to the UV SFRs assuming that \lya\ is absorbed by 24\% on average. 
The dashed line is the best linear fit (in logarithmic space) to the 
relatively robust measurements (filled circles). The \lya\ SFRs are 
increasingly smaller than the UV SFRs towards higher SFRs, mostly due to
the selection effect shown in Figure 6. It could also be due partly to
the smaller escape fractions of \lya\ photons in higher-SFR galaxies.}
\end{figure}

\begin{figure}	
\epsscale{1.1}
\plottwo{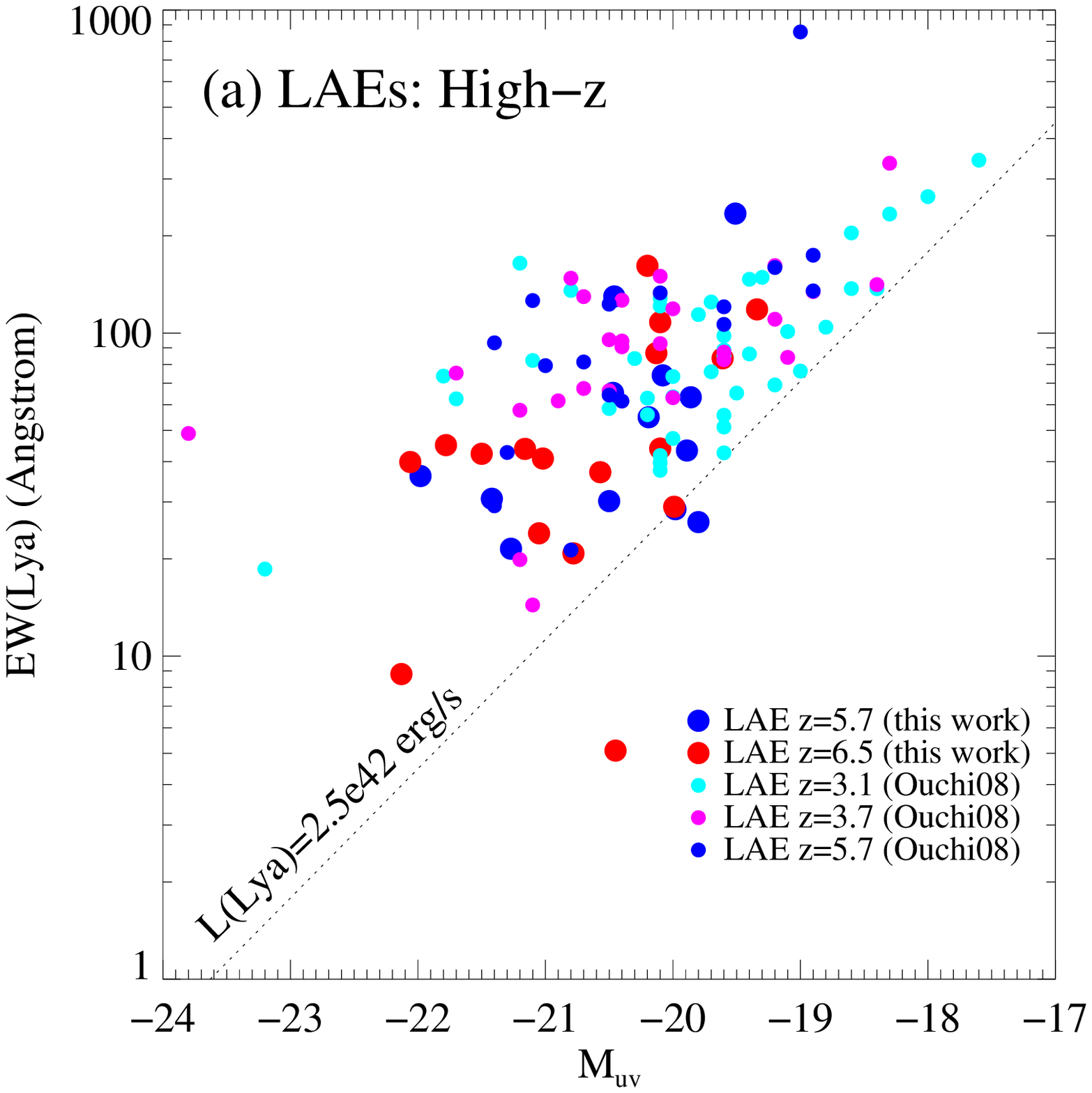}{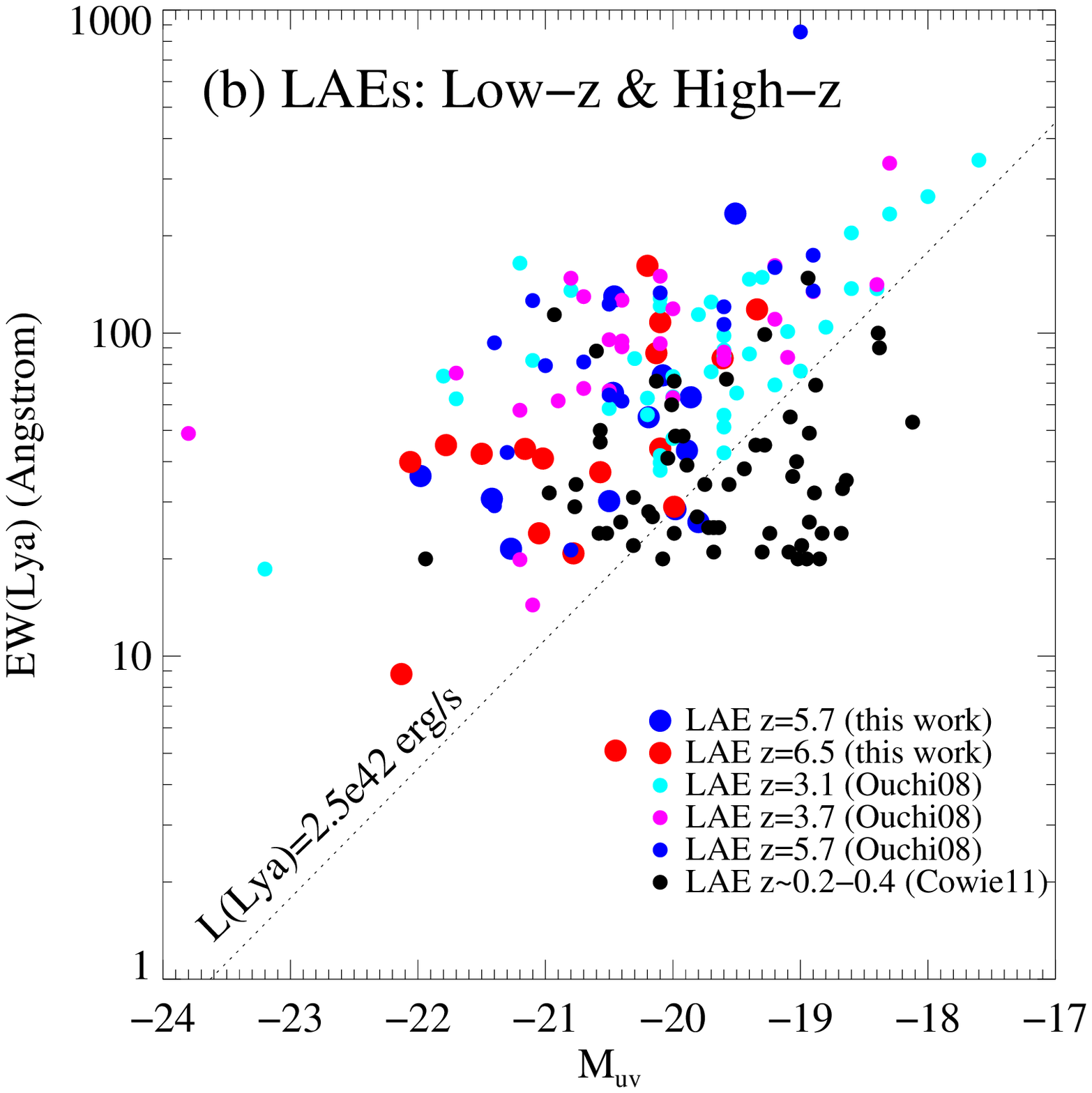}
\caption{EW(\lya) vs. $M_{\rm UV}$ for LAEs. (a) EW(\lya) is plotted as a
function of $M_{\rm UV}$ for the $z\simeq5.7$ and 6.5 LAEs in our sample, 
as well as for the $z\sim3.1$, 3.7, and 5.7 LAEs by \citet{ouc08}. As for
$M_{\rm UV}$, we used $M_{1500}$ listed in Table~2 for our sample.  
For the sample of \citet{ouc08}, we used the published $M_{\rm UV}$, which 
correspond to $M_{1300-1600}$ depending on redshift.  The diagonal dotted line
  is defined by a limiting \lya\ line luminosity of $2.5 \times
  10^{42}$ erg s$^{-1}$, which roughly corresponds to the limits of
  these LAE surveys.  The slope of the EW(\lya)--$M_{\rm UV}$
  correlation is largely shaped by the limiting luminosity.
(b) GALEX-selected $z\simeq0.2-0.4$ LAEs \citep{cowie11b}
  are added for comparison.  $M_{\rm UV}$ was derived from the GALEX near-UV
  photometry in \citet{cowie10}.  The effective wavelength of the near-UV
  band is 2315.7 \AA\ \citep{mor07}, so at redshifts $z\simeq0.2-0.4$, 
the near-UV band is sampling $\sim$1650--1930 \AA\ in the
  rest-frame.  The near-UV, instead of the far-UV band, was used because the
  latter contains the redshifted \lya\ emission. The figure indicates that
the EW(\lya)--$M_{\rm UV}$ relation is largely shaped by the limiting 
luminosity.}
\end{figure}

\clearpage
\begin{figure}	
\epsscale{0.7}
\plotone{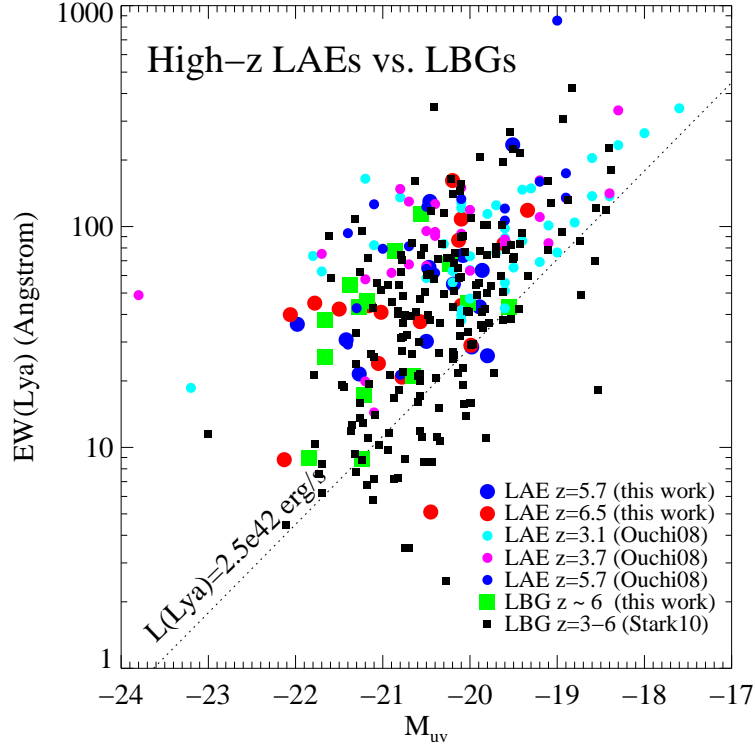}
\caption{Same as Figure~8, but for LBGs from this work
  ($z\simeq6$) and from \citet{sta10} ($z\simeq3-6$).  A number of LBGs from
  \citet{sta10} fall below the L(\lya)$=2.5\times10^{42}$ erg
  s$^{-1}$ line, reflecting the deeper limiting \lya\ line flux
  of the data.  The overall distribution of the points appears similar
  between the LAE and LBG samples.}
\end{figure}

\begin{figure} 
\epsscale{0.7}
\plotone{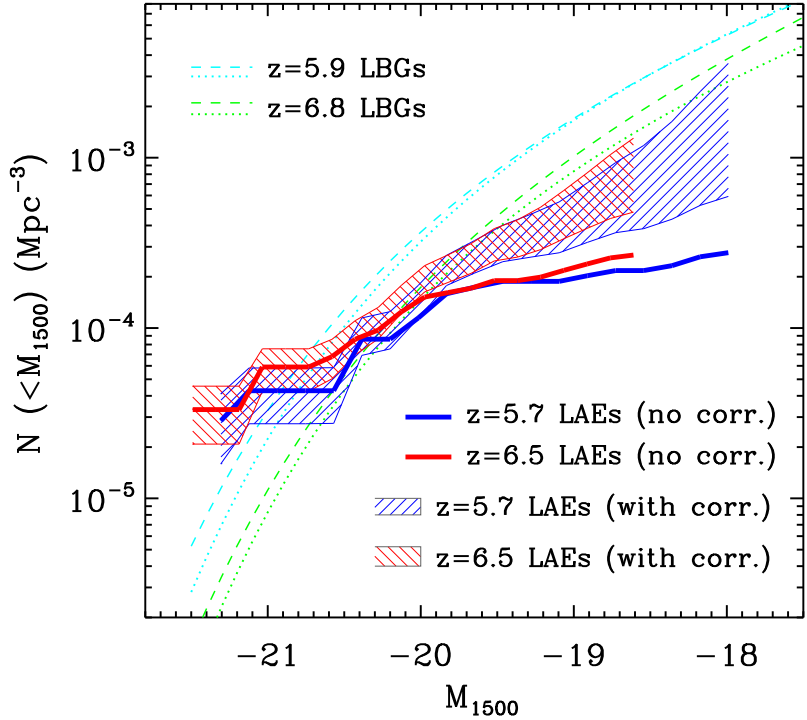}
\caption{UV continuum LFs of LAEs with $\rm EW>20$ \AA\ at $z\simeq5.7$ and 
6.5. The thick blue and red lines represent the spatial densities of the LAEs 
in our sample, without incompleteness correction to $\rm EW=20$ \AA. 
The shaded regions show the UV LFs of LAEs corrected to EW(Ly$\alpha$)$=$ 20 
\AA. Their lower and upper boundaries represent the lower and upper limits of 
the LFs. The $1\sigma$ statistical uncertainties have been included.
The cyan and green curves represent the UV LFs of LBGs at $z\simeq5.9$ 
and 6.8 \citep{bou07,bou11,mcl09,ouc09}. Compared to LBGs, LAEs have 
comparable spatial densities in the bright end, but significantly lower 
densities in the faint end.}
\end{figure}

\clearpage
\begin{figure} 
\epsscale{0.7}
\plotone{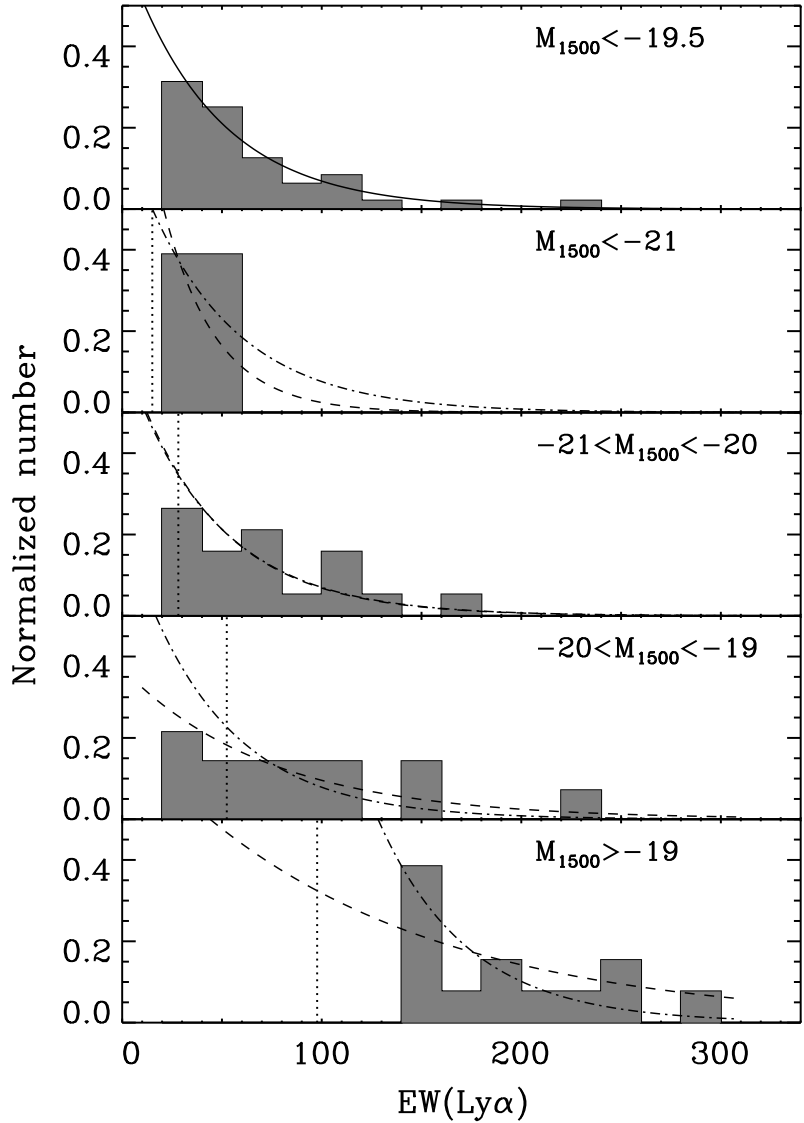}
\caption{Normalized \lya\ EW distribution of the LAEs in our sample.
The vertical dotted lines represent the approximate EW limits for complete LAE 
samples at the given $M_{1500}$ ranges. We use an exponential function to fit 
the EW distribution. We adopt two extreme $e$-folding widths ($b=0.225$ and 0) 
in Section 5.3.  The solid curve in the top panel is the best fit to the LAEs 
at $M_{1500}<-19.5$ mag (it is the same for the two cases).  
The dashed ($b=0.225$) and dash-dotted ($b=0$) curves in the other panels are 
the predicted EW distributions for the two extreme cases. The real 
distribution should be between the two.}
\end{figure}

\clearpage
\begin{figure}
\epsscale{0.9}
\plotone{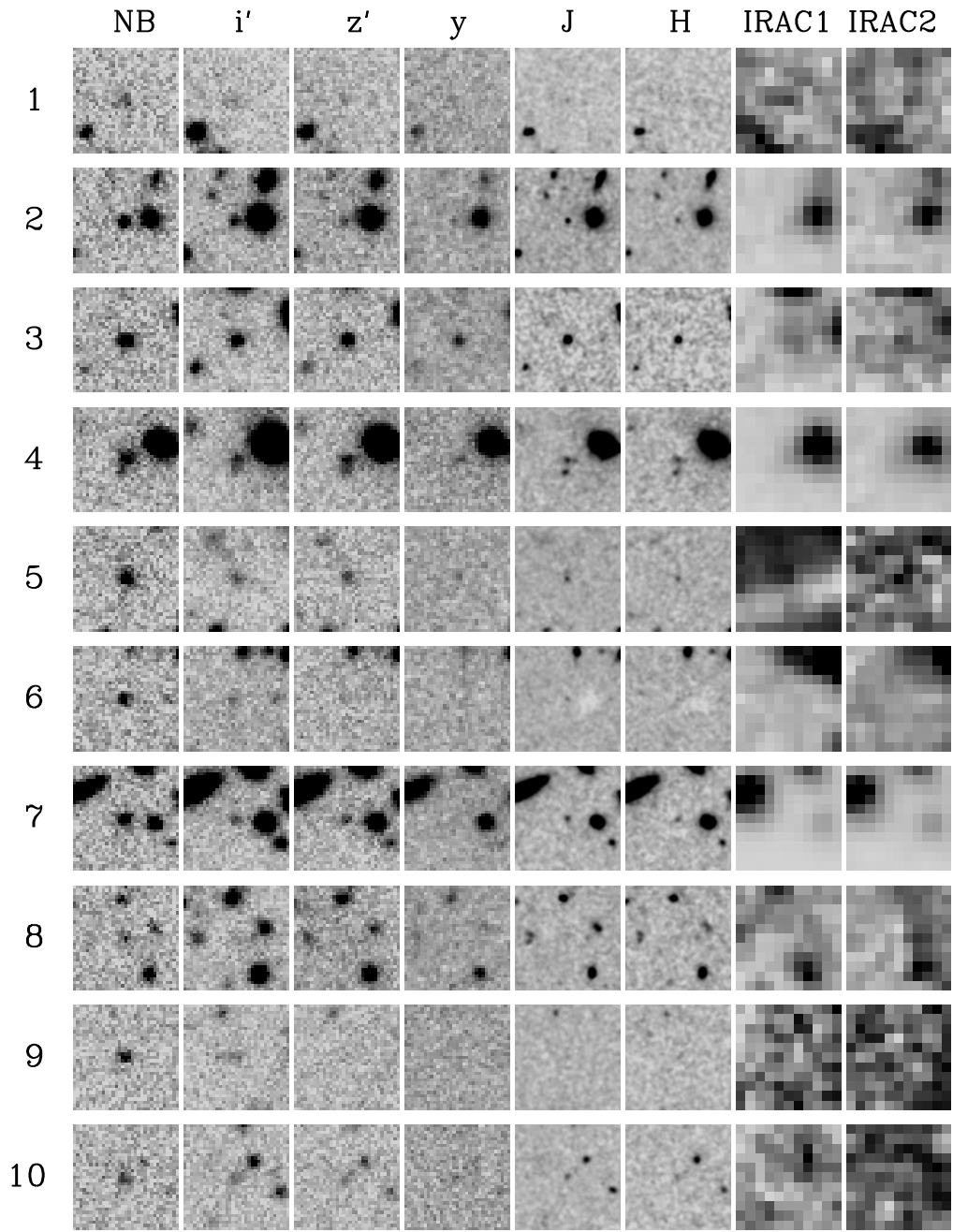}
\caption{Thumbnail images of the galaxies in our sample. 
The galaxies are in the middle of 
the images. The size of the images is $6\farcs6 \times 6\farcs6$ (north is up 
and east to the left). An image is blank if the data is either not available 
or meaningless in this band.}
\end{figure}

\clearpage
\addtocounter{figure}{-1}
\begin{figure}
\plotone{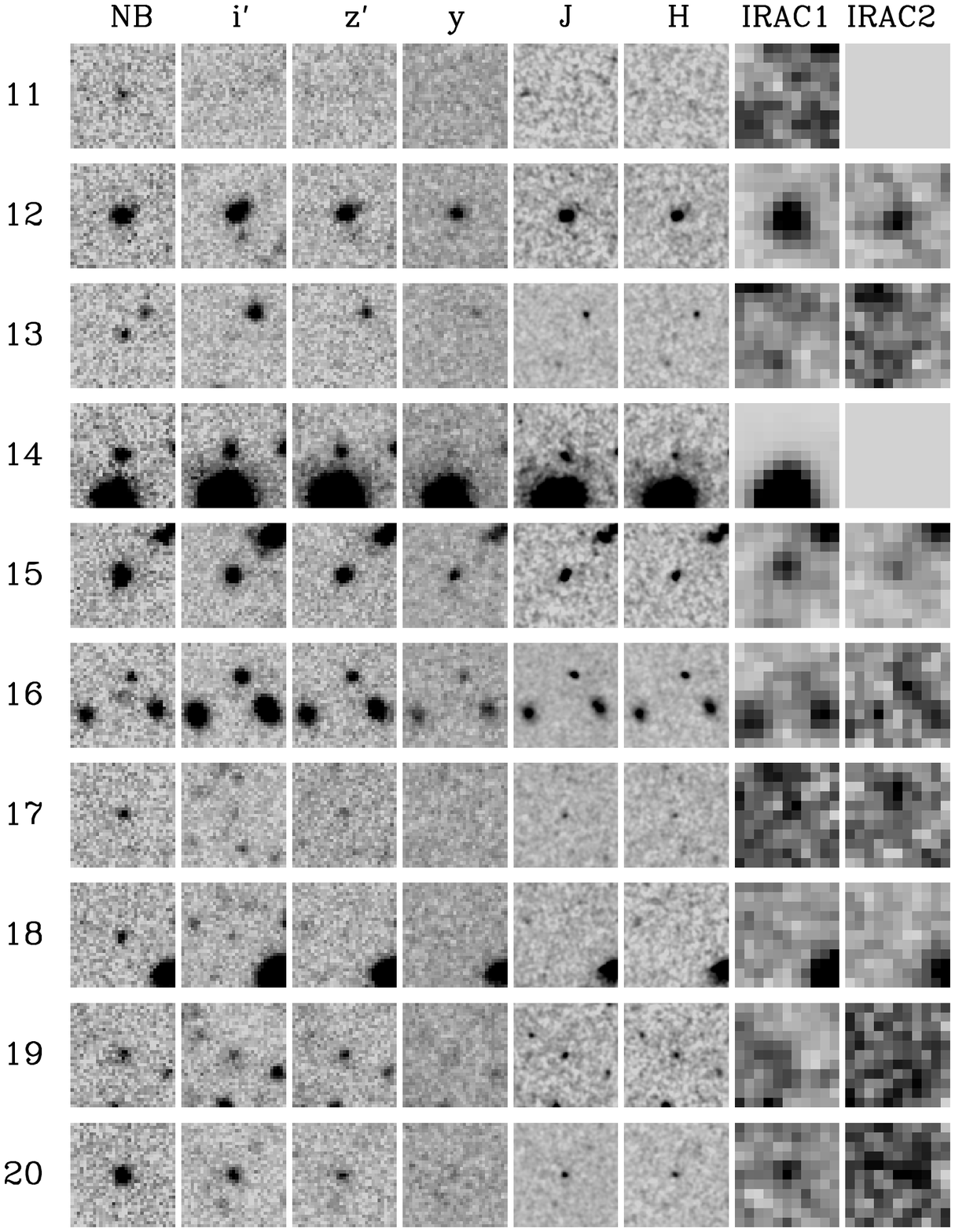}
\caption{Continued.}
\end{figure}

\clearpage
\addtocounter{figure}{-1}
\begin{figure}
\plotone{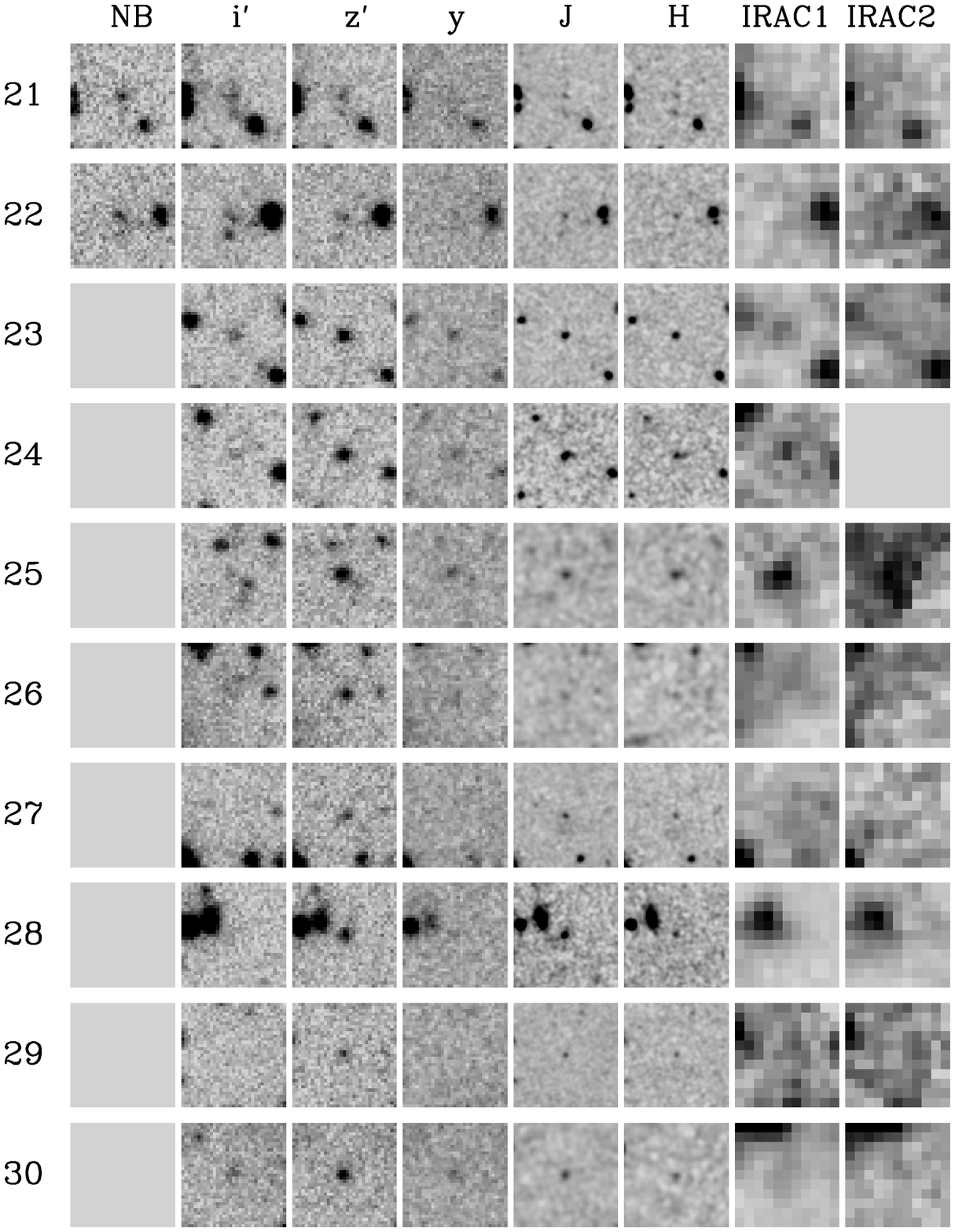}
\caption{Continued.}
\end{figure}

\clearpage
\addtocounter{figure}{-1}
\begin{figure}
\plotone{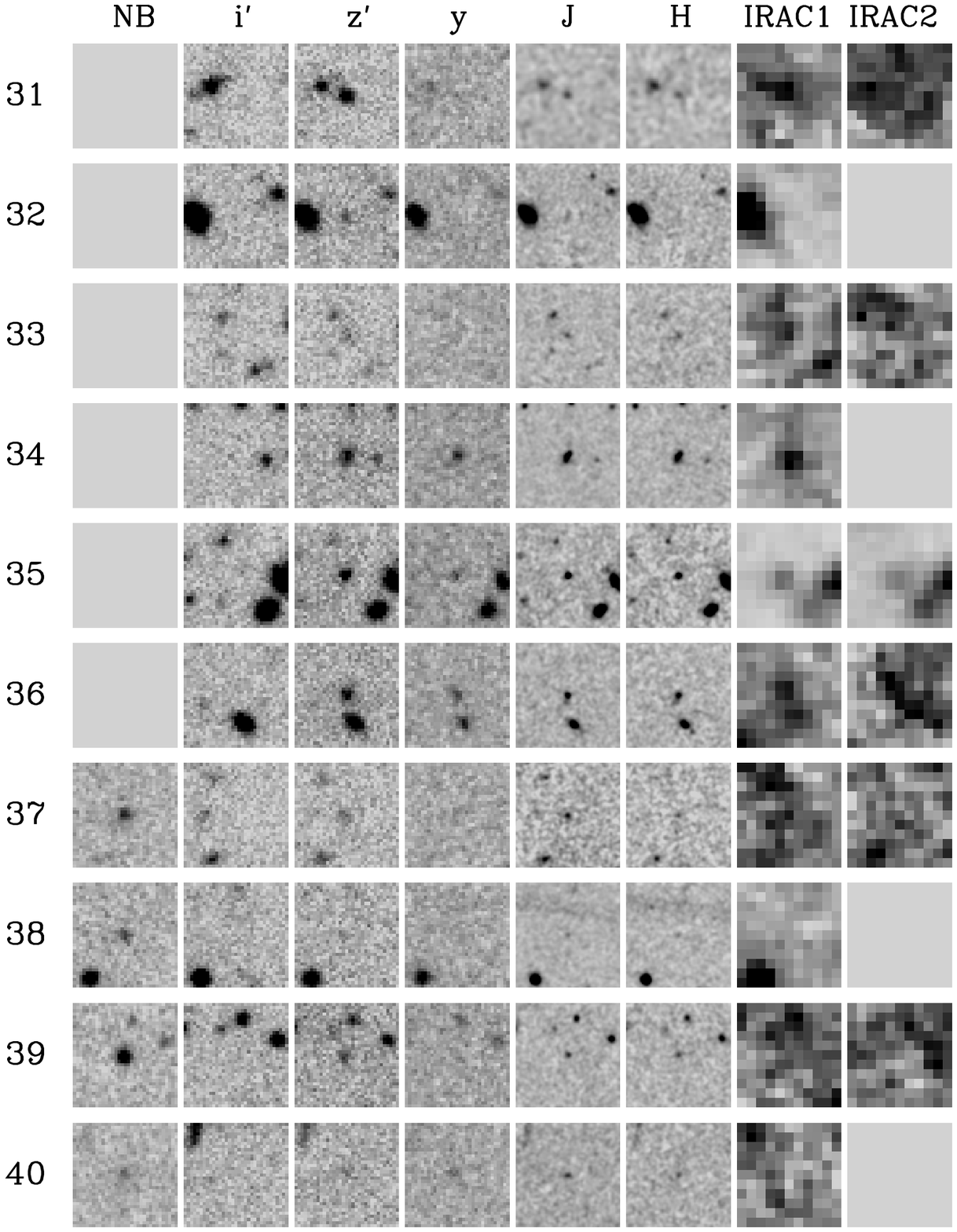}
\caption{Continued.}
\end{figure}

\clearpage
\addtocounter{figure}{-1}
\begin{figure}
\plotone{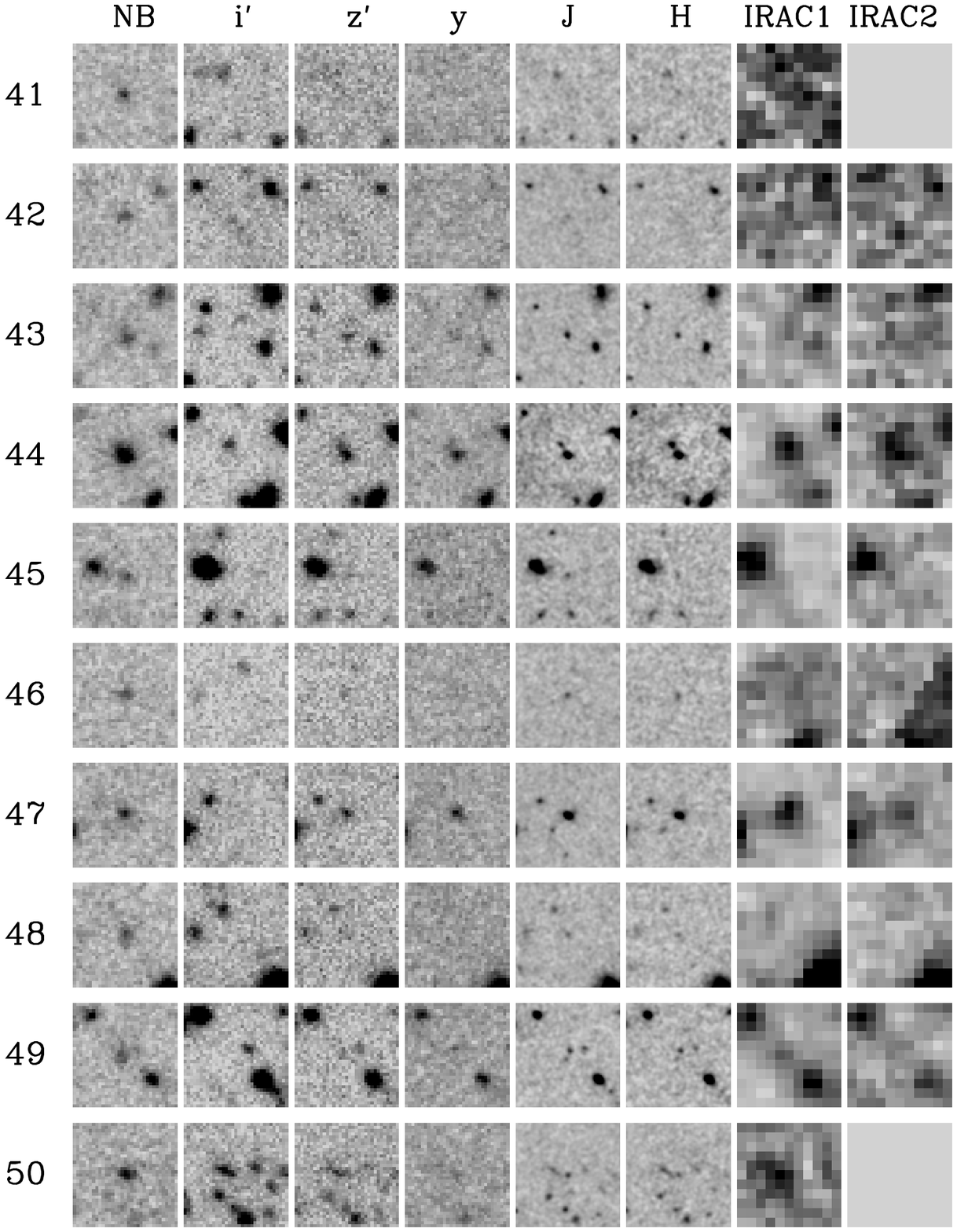}
\caption{Continued.}
\end{figure}

\clearpage
\addtocounter{figure}{-1}
\begin{figure}
\plotone{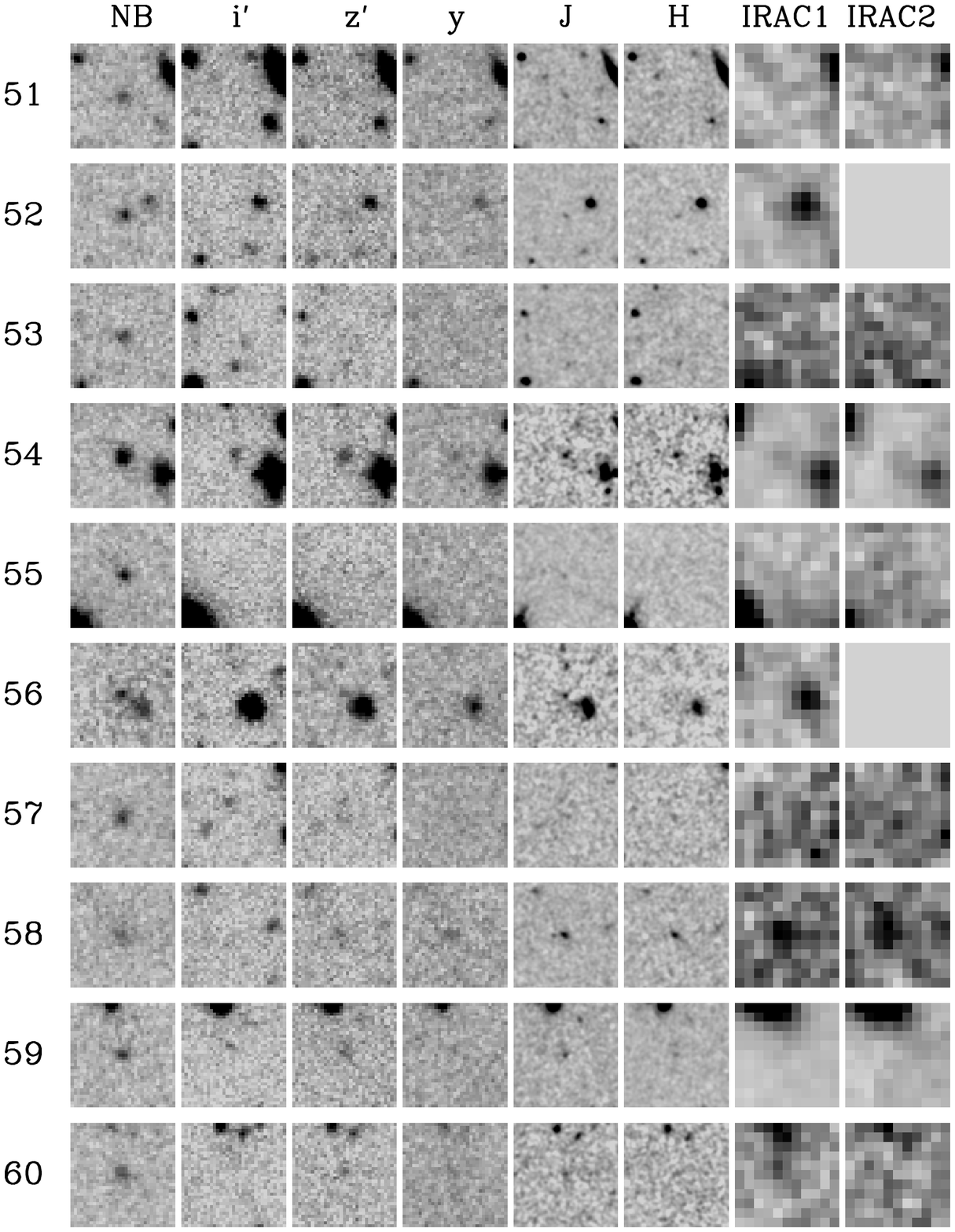}
\caption{Continued.}
\end{figure}

\clearpage
\addtocounter{figure}{-1}
\begin{figure}
\plotone{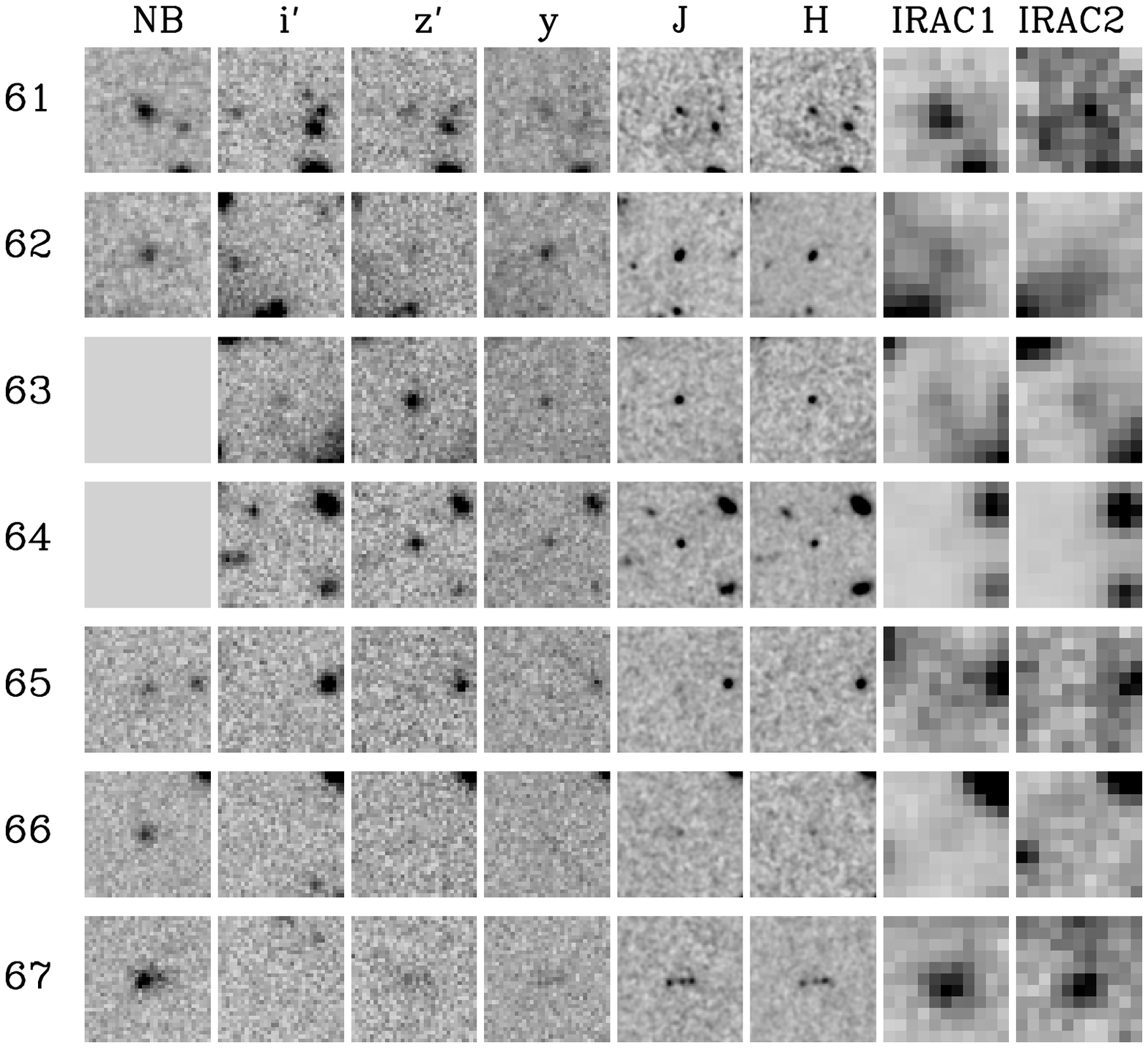}
\caption{Continued.}
\end{figure}

\end{document}